\documentclass[aps,twocolumn]{revtex4}
\usepackage{epsf}
 
\begin{document}
      \title{Flow equation approach to the linear response theory 
             of superconductors}

     \author{M.\ Zapalska and T.\ Doma\'nski}
\affiliation{
             Institute of Physics, 
	     M.\ Curie Sk\l odowska University, 
             20-031 Lublin, Poland} 
      \date{\today}

\begin{abstract}
We apply the flow equation method for studying the current-current
response function of electron systems with the pairing instability. 
To illustrate the specific scheme in which the flow equation procedure 
determines the two-particle Green's functions we reproduce the standard 
response kernel of the BCS superconductor. We next generalize this 
non-perturbative treatment considering the pairing field fluctuations. 
Our study indicates  that the residual diamagnetic behavior detected 
above the transition temperature in the cuprate superconductors 
can originate from the noncondensed preformed pairs.
\end{abstract}  


\maketitle

\section{Introduction}

Non-perturbative scheme of the flow equation method introduced 
by F.\ Wegner \cite{Wegner-94} and independently by K.\ Wilson 
with S.\ G\l azek \cite{Wilson-94} proved to be a useful tool 
for investigating a number of problems in the condensed matter 
physics \cite{Kehrein_book}, mesoscopic systems \cite{Uhrig-10}, 
quantum optics \cite{Kriel} and quantum chromodynamics 
\cite{Gubankova-00}. This procedure has been also recently 
applied to study the non-equilibrium transport phenomena 
of the correlated nanoscopic systems \cite{Kehrein_dots}.
The main idea is rather simple and relies on a continuous 
process which, step by step, transforms the Hamiltonian  
to a diagonal or at least block-diagonal structure. One 
can use for this purpose various operators, depending on 
the subtleties of the discussed problem \cite{Kehrein_book}.  

Such continuous diagonalization scheme is reminiscent of 
the Renormalization Group (RG) technique \cite{Wilson_NRG}. 
They are similar with regard to the treatment of high/low 
energy states (fast/slow modes). In initial steps of 
the continuous transformation procedure mainly the most 
off-diagonal terms (i.e.\ high energy sector) are dealt 
with. Subsequently, the remaining parts closer to the 
diagonal are transformed. Since different energy scales 
are successively transformed/renormalized  one by one  
the algorithm of flow equation method is relative to 
the family of RG formulations \cite{Metzner-11}. 
Let us remark that such techniques are in principle 
unrestricted by limitations of the usual perturbative 
methods.

In this paper we: (1) formulate the current-current response
function for the superconducting system using the flow equation 
method, and (2) extend such scheme to a state of the preformed 
pairs which above the critical temperature $T_{c}$ loose the 
long-range coherence. Our study is motivated by the recent 
torque magnetometry data of the Princeton group \cite{Ong-10} 
revealing the diamagnetic features well preserved above 
$T_{c}$ in the lanthanum and yttrium cuprate oxides. 
Similar indications have been also reported from 
the $dc$ susceptibility measurements 
for Bi$_{2.2}$Sr$_{1.8}$Ca$_{2}$Cu$_{3}$O$_{10+\delta}$
\cite{Iye-10} and from the high-resolution SQUID data
for Sm-based underdoped YBCO compounds \cite{Bernardi-10}. 
Since the observed diamagnetic response is rather strong
it can be hardly assigned to the Ginzburg-Landau 
fluctuations \cite{Larkin-Varlamov}. 

Adopting argumentation discussed in the literature on 
the microscopic \cite{Ranninger-85,Friedberg-89,Micnas-90,
Auerbach-02,Rice-09,Ranninger-10} and the phenomenological 
grounds \cite{Geshkenbein-97,Senthil-09} we consider 
the system consisting of the preformed local pairs (of 
arbitrary origin) coexisting and interacting with the 
itinerant electrons. Using the flow equation approach 
we analyze the diamagnetism within such scenario. Our 
study can be regarded as complementary to the recent 
Quantum Monte  Carlo (QMC) simulations for the same 
cooperon-fermion model \cite{Troyer-11}. It has also 
some resemblance to considerations of the superconducting 
fluctuations beyond the gaussian approximation carried out 
in the t-J model \cite{Castellani-11}.

We start by discussing the usual BCS model, treating it 
as a testing field for formulation of the linear response 
theory in terms of the flow equation method (readers less
interested in the methodological details can skip this 
section). We next  apply the same treatment to the mixture 
of electrons interacting through the Andreev scattering 
with the preformed local pairs. We determine the 
current-current response function and try to asses 
the diamagnetic response above $T_{c}$. In summary, we 
point out the main conclusions and give a list of problems 
which might be of interest for further studies.

\section{BCS superconductor} 

Let us first briefly illustrate how the flow equation procedure 
determines the quasiparticle spectrum and the corresponding
response function for the usual BCS model
\begin{eqnarray}
\hat{H} & = & \sum_{{\bf k},\sigma} \xi_{\bf k} 
\hat{c}_{{\bf k}\sigma}^{\dagger} \hat{c}_{{\bf k}\sigma} 
- \sum_{\bf k} \left( \Delta_{\bf k}  
\hat{c}_{{\bf k}\uparrow}^{\dagger} \hat{c}_{-{\bf k}
\downarrow}^{\dagger} + h.c. \right) 
\label{hamil}
\end{eqnarray}
describing electrons coupled to the pairing field 
$\Delta_{\bf k}$. We use here the standard notation 
for the creation (annihilation) operators  
$\hat{c}_{{\bf k}\sigma}^{\dagger}$ 
($\hat{c}_{{\bf k}\sigma}$) and denote by $\xi_{\bf k}
=\varepsilon_{\bf k}\!-\!\mu$ the energies 
measured with respect to the chemical potential $\mu$. 
Formally, $\Delta_{\bf k}$ can be thought as the 
Bose-Einstein condensate of the Cooper pairs 
$\Delta_{\bf k}=-\sum_{\bf q} V_{{\bf k},{\bf q}} 
\langle \hat{c}_{-{\bf q}\downarrow}\hat{c}_{{\bf q}
\uparrow}\rangle$ which are formed by some attractive 
potential  $V_{{\bf k},{\bf q}}<0$.

\subsection{The continuous diagonalization}

The continuous diagonalization of the reduced BCS Hamiltonian 
(\ref{hamil}) has been considered in the original paper 
by F.\ Wegner \cite{Wegner-94} and by several other authors 
\cite{Stelter,Domanski-07}. We briefly recollect the main 
steps of such procedure (see appendix A for the procedural
details) which shall be useful for formulating the linear 
response theory. 

Following Wegner \cite{Wegner-94} we choose the generating 
operator $\hat{\eta}(l)$ defined by Eq.\ (\ref{wegner}), 
which for the BCS model (\ref{hamil}) has the following 
structure 
\begin{eqnarray}                                                 
\hat{\eta}(l) = 2 \sum_{\bf k} \xi_{\bf k}(l) \left(
\Delta_{\bf k}(l) \hat{c}_{{\bf k}\uparrow}^{\dagger} 
\hat{c}_{-{\bf k}\downarrow}^{\dagger} -\Delta^{*}
_{\bf k}(l) \hat{c}_{-{\bf k}\downarrow}
\hat{c}_{{\bf k}\uparrow} \right) .
\label{eta}
\end{eqnarray}
Transformation of the Hamiltonian $\hat{H}(l)$ proceeds
as long as $ \eta(l)$ is finite, which occurs until 
$\Delta_{\bf k}(l) \rightarrow 0$. This is achieved 
in the asymptotic limit $l\rightarrow\infty$ (\ref{limit}). 

Substituting (\ref{eta}) to the general flow equation (\ref{general}) 
for the Hamiltonian (\ref{hamil}) we obtain \cite{Domanski-07}
\begin{eqnarray}
\frac{d}{dl}\; \mbox{\rm ln} \xi_{\bf k}(l) & = & 
4 |\Delta_{\bf k}(l)|^{2} \label{xi_flow} ,
\\
\frac{d}{dl}\; \mbox{\rm ln} \Delta_{\bf k}(l)  & = & 
-4 (\xi_{\bf k}(l))^{2} .
\label{Delta_flow} 
\end{eqnarray} 
These equations (\ref{xi_flow},\ref{Delta_flow}) 
yield an exponential {\em flow} 
\begin{eqnarray} 
 \Delta_{\bf k}(l) = \Delta_{\bf k} \; e^{-4\int_{0}^{l} dl'
[\xi_{\bf k}(l')]^2}
\end{eqnarray}
and $\xi_{\bf k}(l) = \xi_{\bf k} e^{4\int_{0}^{l} dl'
|\Delta_{\bf k}(l')|^2}$, therefore the off-diagonal 
term $\Delta_{\bf k}(l)$ vanishes in the limit
$l\rightarrow\infty$. Combining (\ref{xi_flow},
\ref{Delta_flow}) we moreover notice that  
$\frac{d}{dl} \left\{ \xi_{\bf k}^{2}(l)) + 
|\Delta_{\bf k}(l)|^{2} \right\} = 0$ which
implies the invariance $\xi_{\bf k}^{2}(l) 
+ |\Delta_{\bf k}(l)|^{2}  = {\mbox{\rm const}}$. Due to 
$\Delta_{\bf k}(\infty)=0$ we conclude that 
the quasiparticle energies take the following BCS form 
\begin{eqnarray}
\tilde{\xi}_{\bf k} = \mbox{sgn}({\xi_{\bf k}}) 
\sqrt{\xi^{2}_{\bf k} + |\Delta_{\bf k}|^{2} } ,
\label{bogol}
\end{eqnarray} 
where we introduced the shorthand notation for the
asymptotic value $\tilde{\xi}_{\bf k} \equiv 
\lim_{l\rightarrow \infty}\xi_{\bf k}(l)$.

\subsection{The single particle excitations}

As an illustration how one can use this procedure to obtain
various Green's functions let us derive the single particle 
excitation spectrum determined by $G_{\sigma}({\bf k},\tau)
=-\hat{T}_{\tau} \langle \hat{c}_{{\bf k} \sigma}(\tau) 
\hat{c}_{{\bf k} \sigma}^{\dagger}\rangle$, where $\hat{T}
_{\tau}$ denotes chronological ordering and $\tau$ stands 
for the imaginary time. Applying 
(\ref{eta}) to the flow equation (\ref{O_flow}) 
for the creation and annihilation operators we infer 
the Bogoliubov ansatz 
\begin{eqnarray}
\hat{c}_{{\bf k}\uparrow}(l) &=& u_{\bf k}(l) \hat{c}_{{\bf k}
\uparrow}+v_{\bf k}(l) \hat{c}_{-{\bf k}\downarrow}^{\dagger} ,
\label{Ansatz1} \\
\hat{c}^{\dagger}_{-{\bf k}\downarrow}(l) &=&  
- v_{\bf k}(l) \hat{c}_{{\bf k}\uparrow} + 
u_{\bf k}(l) \hat{c}_{-{\bf k}\downarrow}^{\dagger}
\label{Ansatz2}
\end{eqnarray}
with the initial boundary conditions $u_{\bf k}(0)\!=\!1$, 
$v_{\bf k}(0)\!=\!0$. Arranging the $l$-dependent coefficients 
if front of $\hat{c}_{{\bf k}\uparrow}$ and 
$\hat{c}_{-{\bf k}\downarrow}^{\dagger}$ on both sides 
of the flow equation (\ref{O_flow}) we find that 
\begin{eqnarray}
\frac{du_{\bf k}(l)}{dl} & = & -2 \xi_{\bf k}(l)
\Delta_{\bf k}^{*}(l)  v_{\bf k}(l),
\label{u_flow} \\
\frac{dv_{\bf k}(l)}{dl}  & = & 2 \xi_{\bf k}(l)
\Delta_{\bf k}(l) u_{\bf k}(l) .
\label{v_flow}
\end{eqnarray}  
From the equations (\ref{u_flow},\ref{v_flow}) we can see that 
the sum rule $u_{\bf k}^{2}(l) +v_{\bf k}^{2}(l)\!=\!1$ is properly 
conserved. To determine the needed asymptotic values 
we can rewrite (\ref{u_flow}) as $\frac{du_{\bf k}(l)}
{v_{\bf k}(l)}\!=\!-2 \xi_{\bf k}(l)\Delta_{\bf k}(l) dl$ 
and substituting $v_{\bf k}(l)=\sqrt{1-u^{2}_{\bf k}(l)}$ 
we can analytically solve the integral $\int_{0}^{\infty} dl$. 
In the asymptotics we obtain the usual Bogoliubov-Valatin 
coefficients 
\begin{eqnarray} 
\tilde{u}_{\bf k}^{2} = 1 -  \tilde{v}_{\bf k}^{2}
= \frac{1}{2} \left[ 1 + \frac{\xi_{\bf k}}
{\tilde{\xi}_{\bf k}} \right]  . 
\end{eqnarray}    
Fourier transform of the single particle Green's function
\begin{eqnarray} 
G_{\sigma}({\bf k},i\omega) = \beta^{-1}
\int_{0}^{\beta} d\tau e^{-i\omega\tau} 
G_{\sigma}({\bf k},\tau)
\label{Fourier}
\end{eqnarray} 
(where $\beta^{-1}\!=\!k_{B}T$) takes hence the two-pole structure
\begin{eqnarray}
G_{\sigma}({\bf k},i\omega) =
\frac{\tilde{u}_{\bf k}^{2}}{i\omega-\tilde{\xi}_{\bf k}}
+\frac{\tilde{v}_{\bf k}^{2}}{i\omega+\tilde{\xi}_{\bf k}}
\end{eqnarray}
signaling the particle-hole mixing, characteristic  
for the BCS state.

\subsection{The linear response theory} 

We now adopt the same procedure for studying a response of the BCS
superconductor to a weak electromagnetic field ${\bf A}({\bf r},t)$. 
In the linear response the induced current ${\bf J}({\bf r},t)$ is 
assumed to be proportional to the perturbation, i.e.\ 
${\bf J}({\bf r},t)=-\int d{\bf r}' \int_{-\infty}^{t} 
dt' K({\bf r}\!-\!{\bf r}',t\!-\!t') {\bf A}({\bf r}',t')$. 
Fourier transform of the kernel function consists of the 
diamagnetic and paramagnetic contributions \cite{Fetter_book}
\begin{eqnarray}
K_{\alpha,\beta}({\bf q},\omega) = \frac{ne^2}{m} \; 
\delta_{\alpha,\beta}  +e^{2} \; \Pi_{\alpha,\beta}
({\bf q},\omega) .
\label{linear}
\end{eqnarray} 
From now onwards, by $\alpha$, $\beta$ we shall denote the 
Cartesian coordinates $x$, $y$, and $z$. The paramagnetic 
term can be expressed by the (analytically continued) 
Fourier transform (\ref{Fourier}) of the current-current 
Green's function 
\begin{eqnarray} 
\Pi_{\alpha,\beta}({\bf q},\tau) \equiv -  \;
\langle \hat{T}_{\tau} \hat{j}_{\bf {q},\alpha}(\tau) 
\hat{j}_{-{\bf q},\beta} \rangle ,
\label{jj}
\end{eqnarray} 
where the corresponding current operator $\hat{\bf j}_{{\bf q}}=
\hat{\bf j}^{\uparrow}_{{\bf q}}+\hat{\bf j}^{\downarrow}_{{\bf q}}$ 
consists of the spin $\uparrow$ and $\downarrow$ contributions
\begin{eqnarray} 
\hat{\bf j}^{\sigma}_{\bf {q}} &=& \sum_{\bf k} {\bf v}_{{\bf k}
+\frac{\bf q}{2}} \hat{c}_{{\bf k}, \sigma}^{\dagger} 
\hat{c}_{{\bf k}+{\bf q}, \sigma} 
\label{j_def}
\end{eqnarray}
with the velocity ${\bf v}_{\bf k}\!=\!\hbar^{-1}
\nabla_{\bf k}\varepsilon_{\bf k}$.

The standard way for computing the current-current response 
function (\ref{jj}) is based on the diagrammatic bubble-type
contributions involving the particle and hole propagators 
and another contribution  from the off-diagonal (in Nambu 
notation) single-particle propagators. In this section 
we retrieve the standard BCS result \cite{Fetter_book} 
using the flow equation routine.

To guess the relevant {\em flow} of the current
operators we start by analyzing the initial $(l\!=\!0)$ 
derivative
\begin{eqnarray}
\left( \frac{d \; \hat{\bf j}^{\sigma}_{{\bf q}}(l)}{dl} \right)_{l=0} 
= \left[ \hat{\eta}(l), \hat{\bf j}^{\sigma}_{{\bf q}}(l) 
\right]_{l=0}
\label{flow_j_init}
\end{eqnarray}
where $\hat{\bf j}^{\sigma}_{{\bf q}}(l\!=\!0)$ corresponds 
to the definition (\ref{j_def}). Using the generating operator 
(\ref{eta}) we find that
\begin{widetext}
\begin{eqnarray}
\left( \frac{d \; \hat{\bf j}^{\sigma}_{{\bf q}}(l)}{dl} \right)_{l=0} 
&=& \pm 2\sum_{\bf k} {\bf v}_{{\bf k}+\frac{\bf q}{2}} 
\left ( \xi_{\bf k} \Delta_{\bf k} \hat{c}_{-{\bf k},
-\sigma} \hat{c}_{{\bf k}+{\bf q},\sigma} + \xi_{{\bf k}+{\bf q}} 
\Delta_{{\bf k}+{\bf q}} \hat{c}_{{\bf k},\sigma}^{\dagger} 
\hat{c}_{-({\bf k}+{\bf q}),-\sigma}^{\dagger} \right ) ,
\label{j_init} 
\end{eqnarray}
where the sign $+$ ($-$) refers to spin $\uparrow$ ($\downarrow$).
Eq.\ (\ref{j_init}) unambiguously implies the following 
$l$-dependent parametrization 
\begin{eqnarray} 
\begin{array}{lcl}
\hat{\bf j}^{\uparrow}_{{\bf q}}(l) &=& \sum_{\bf k} 
{\bf v}_{{\bf k}+\frac{\bf q}{2}} \left( {\cal{A}}_{\bf k,q}(l) 
\hat{c}_{{\bf k},\uparrow}^{\dagger}\hat{c}_{{\bf k}+{\bf q},\uparrow} 
+{\cal{B}}_{\bf k,q}(l) \hat{c}_{-{\bf k},\downarrow} 
\hat{c}_{-({\bf k}+{\bf q}),\downarrow}^{\dagger}   
+   {\cal{D}}_{\bf k,q}(l)  \hat{c}_{{\bf k},\uparrow}^{\dagger} 
\hat{c}_{-({\bf k}+{\bf q}),\downarrow}^{\dagger} + {\cal{F}}_{\bf k,q}(l) 
\hat{c}_{-{\bf k},\downarrow} \hat{c}_{{\bf k}+{\bf q},\uparrow} 
 \right)  \\
\hat{\bf j}^{\downarrow}_{{\bf q}}(l) &=& \sum_{\bf k} 
{\bf v}_{{\bf k}+\frac{\bf q}{2}} \left( {\cal{A}}_{\bf k,q}(l) 
\hat{c}_{{\bf k},\downarrow}^{\dagger}\hat{c}_{{\bf k}+{\bf q},\downarrow} 
+{\cal{B}}_{\bf k,q}(l) \hat{c}_{-{\bf k},\uparrow} 
\hat{c}_{-({\bf k}+{\bf q}),\uparrow}^{\dagger}   
-   {\cal{F}}_{\bf k,q}(l)  \hat{c}_{{\bf k},\uparrow}^{\dagger} 
\hat{c}_{-({\bf k}+{\bf q}),\downarrow}^{\dagger} - {\cal{D}}_{\bf k,q}(l) 
\hat{c}_{-{\bf k},\downarrow} \hat{c}_{{\bf k}+{\bf q},\uparrow} 
 \right) \end{array}
\label{j_ansatz} 
\end{eqnarray}
with the initial boundary conditions ${\cal{A}}_{\bf k,q}(0)\!=\!1$ 
and  ${\cal{B}}_{\bf k,q}(0)\!=\!{\cal{D}}_{\bf k,q}(0)\!=\!
{\cal{F}}_{\bf k,q}(0)\!=\!0$.  All $l$-dependent coefficients have 
to be determined applying the ansatz (\ref{j_ansatz}) in the flow 
equation (\ref{O_flow}) for the current operators 
$\hat{\bf j}^{\sigma}_{\bf q}(l)$. 
On this basis we obtain the following set of equations
\begin{eqnarray}
\frac{d {\cal{A}}_{\bf k,q}(l)}{dl}  &=& -2\left[ 
\xi_{\bf k+q}(l)\Delta_{\bf k+q}(l) {\cal{D}}_{\bf k,q}(l)
+ \xi_{\bf k}(l)\Delta_{\bf k}(l) {\cal{F}}_{\bf k,q}(l) \right], 
\label{flow_alpha}\\
\frac{d {\cal{B}}_{\bf k,q}(l)}{dl}  &=&2\left[
\xi_{\bf k}(l)\Delta_{\bf k}(l){\cal{D}}_{\bf k,q}(l)   
+\xi_{\bf k+q}(l)\Delta_{\bf k+q}(l) {\cal{F}}_{\bf k,q}(l) \right],
\label{flow_gamma} \\
\frac{d {\cal{D}}_{\bf k,q}(l)}{dl}  &=& 2 \left[
\xi_{\bf k+q}(l)\Delta_{\bf k+q}(l) {\cal{A}}_{\bf k,q}(l)
- \xi_{\bf k}(l)\Delta_{\bf k}(l){\cal{B}}_{\bf k,q}(l) \right] ,
\label{flow_beta}\\
\frac{d {\cal{F}}_{\bf k,q}(l)}{dl}  &=& 2 \left[
\xi_{\bf k}(l)\Delta_{\bf k}(l) {\cal{A}}_{\bf k,q}(l) 
- \xi_{\bf k+q}(l)\Delta_{\bf k+q}(l) {\cal{B}}_{\bf k,q}(l) \right].
\label{flow_eta}
\end{eqnarray}
By inspecting (\ref{flow_alpha}-\ref{flow_eta}) we can notice that
\begin{eqnarray}
\frac{d}{dl} \left[ {\cal{A}}_{\bf k,q}(l) + {\cal{B}}_{\bf k,q}(l) 
\right]&=&2\left[ \xi_{\bf k+q}(l)\Delta_{\bf k+q}(l)-\xi_{\bf k}(l)
\Delta_{\bf k}(l) \right] \; \left[ {\cal{F}}_{\bf k,q}(l) 
- {\cal{D}}_{\bf k,q}(l)\right] , \label{id_1} \\
\frac{d}{dl} \left[ {\cal{D}}_{\bf k,q}(l) - {\cal{F}}_{\bf k,q}(l) 
\right]&=& -2 \left[ \xi_{\bf k+q}(l)\Delta_{\bf k+q}(l)-\xi_{\bf k}(l)
\Delta_{\bf k}(l) \right] \; \left[ {\cal{A}}_{\bf k,q}(l) + 
{\cal{B}}_{\bf k,q}(l)\right] ,
\label{id_2}
\end{eqnarray}
which implies
\begin{eqnarray}
\frac{d}{dl} \left[  {\cal{A}}_{\bf k,q}(l) + {\cal{B}}_{\bf k,q}(l) 
\right]^{2} + \frac{d}{dl} \left[ {\cal{D}}_{\bf k,q}(l) - 
{\cal{F}}_{\bf k,q}(l) \right]^{2}  = 0 .
\label{ageb}
\end{eqnarray}
Taking into account the initial boundary conditions we hence obtain 
the following invariance 
\begin{eqnarray}
\left[ {\cal{A}}_{\bf k,q}(l) + {\cal{B}}_{\bf k,q}(l) \right]^{2} 
+ \left[ {\cal{D}}_{\bf k,q}(l) - {\cal{F}}_{\bf k,q}(l) \right]^{2} = 1 
\label{invariance}
\end{eqnarray}
valid for arbitrary $l$, including the limit $l\rightarrow\infty$. 
Combining (\ref{invariance}) with the differential equations 
(\ref{id_1},\ref{id_2}) we exactly determine the asymptotic 
limit values 
\begin{eqnarray}
\left[ \tilde{\cal{A}}_{\bf k,q} + \tilde{\cal{B}}_{\bf k,q} \right]^{2}
&=&  \frac{1}{2} \left( 1 + \frac{\Delta_{\bf k+q}\Delta_{\bf k} + 
\xi_{\bf k+q}\xi_{\bf k}}{\tilde{\xi}_{\bf k+q}\tilde{\xi}_{\bf k}} \right ) ,
\label{coh_1} \\
\left[ \tilde{\cal{D}}_{\bf k,q} - \tilde{\cal{F}}_{\bf k,q} \right]^{2}
&=& \frac{1}{2} \left( 1 - \frac{\Delta_{\bf k+q}\Delta_{\bf k} 
+ \xi_{\bf k+q}\xi_{\bf k}}{\tilde{\xi}_{\bf k+q}\tilde{\xi}_{\bf k}} \right )  ,
\label{coh_2}
\end{eqnarray}
where $\tilde{\xi}_{\bf k}\!=\!\sqrt{\xi^{2}_{\bf k} + 
\Delta^{2}_{\bf k}}$. In the same way we also check that
$\left[  {\cal{A}}_{\bf k,q}(l) - {\cal{B}}_{\bf k,q}(l) \right]^{2} 
- \left[ {\cal{D}}_{\bf k,q}(l) + {\cal{F}}_{\bf k,q}(l) \right]^{2}  
= 1$ thereby the asymptotic values of all coefficients are found 
$\tilde{\cal{A}}_{\bf k,q}\!=\! \tilde{u}_{\bf k}\tilde{u}_{\bf k+q}$,
$\tilde{\cal{B}}_{\bf k,q}\!=\!\tilde{v}_{\bf k}\tilde{v}_{\bf k+q}$,
$\tilde{\cal{D}}_{\bf k,q}\!=\!\tilde{v}_{\bf k}\tilde{u}_{\bf k+q}$,
and $\tilde{\cal{F}}_{\bf k,q}\!=\!\tilde{u}_{\bf k}\tilde{v}_{\bf k+q}$.

Since the transformed Hamiltonian $\hat{H}(\infty)$ is diagonal we can 
easily compute the current-current response function (\ref{jj}) expressing 
it through the particle-hole bubble diagrams (see the left h.s.\ panel in figure 
\ref{plot1}). Finally it is given by
\begin{eqnarray}
\Pi_{\alpha,\beta}({\bf q},i\nu) &=& \sum_{\bf k}
v_{{\bf k}+\frac{\bf q}{2},\alpha} v_{{\bf k}+\frac{\bf q}{2},
\beta}   \left\{ \left[ \tilde{\cal{A}}_{\bf k,q} + 
\tilde{\cal{B}}_{\bf k,q} \right]^{2} 
\left[ f_{FD}(\tilde{\xi}_{\bf k+q}) \!-\! f_{FD}(\tilde{\xi}_{\bf k}) \right] 
\left[ \frac{1}{i \nu \!+\! \tilde{\xi}_{\bf k+q} \!-\! \tilde{\xi}_{\bf k}} 
- \frac{1} {i \nu\! -\! \tilde{\xi}_{\bf k+q}\! +\! \tilde{\xi}_{\bf k}} \right] \right. 
\nonumber \\
&& + \left. \left[ \tilde{\cal{D}}_{\bf k,q} - 
\tilde{\cal{F}}_{\bf k,q} \right]^{2} \left [1 \!-\! 
f_{FD}(\tilde{\xi}_{\bf k+q}) \!-\! f_{FD}(\tilde{\xi}_{\bf k}) \right ] 
\left[ \frac{1}{i \nu \!-\! \tilde{\xi}_{\bf k+q}\!-\! \tilde{\xi}_{\bf k}} 
- \frac{1}{i \nu \!+\! \tilde{\xi}_{\bf k+q}\! +\! \tilde{\xi}_{\bf k}} \right] \right\} ,  
\label{BCS_result}
\end{eqnarray}
where $f_{FD}(\omega)\!=\!\left[\mbox{\rm exp}(\omega/k_{B}T)
\!+\!1\right]^{-1}$ is the Fermi-Dirac distribution function.
We recognize that (\ref{coh_1},\ref{coh_2}) correspond 
to the usual BCS coherence factors $(\tilde{u}_{\bf k+q} 
\tilde{u}_{\bf k} + \tilde{v}_{\bf k+q} \tilde{v}_{\bf k})^{2}$ 
and $(\tilde{u}_{\bf k+q} \tilde{v}_{\bf k} - \tilde{v}_{\bf k+q} 
\tilde{u}_{\bf k})^{2}$ and thereby Eq.\ (\ref{BCS_result}) 
rigorously reproduces the known BCS response function \cite{Fetter_book}.
\end{widetext}

\section{Superconducting fluctuations}

Numerous experimental data \cite{Ong-10,Orenstein-99,Ong-00,
Bergeal-08,Yuli-09,Argonne-08,Villigen-08,Davis-09,
Chatterjee-09,McElroy-08} provided a rather clear 
evidence that the critical temperature $T_{c}$ in the 
underdoped cuprate superconductors  (and similarly in 
the ultracold fermion gasses near the unitary limit 
\cite{Levin-05}) is not related to appearance of 
the fermion pairs but corresponds to the onset 
of their phase coherence. Upon approaching $T_{c}$ 
from above the short-range superconducting correlations 
gradually emerge. 
For instance, the torque magnetometry \cite{Ong-10} 
and other measurements \cite{Iye-10,Bernardi-10}
have detected the diamagnetic properties.

To investigate the electrodynamic properties of 
the non-condensed preformed pairs we consider 
the model 
\begin{eqnarray}
\hat{H} &=& \sum_{{\bf k},\sigma} \xi_{\bf k} 
\hat{c}_{{\bf k}\sigma}^{\dagger} \hat{c}_{{\bf k}\sigma} 
+ \sum_{\bf q} E_{\bf q} 
\hat{b}_{\bf q}^{\dagger} \hat{b}_{\bf q} 
\label{BF} \\
&+& \frac{1}{\sqrt{N}} \; 
\sum_{{\bf k},{\bf p}} g_{{\bf k},{\bf p}} \left(   
\hat{b}_{{\bf k}+{\bf p}}^{\dagger} \hat{c}_{{\bf k}\downarrow}
\hat{c}_{{\bf p}\uparrow} + \hat{b}_{{\bf k}+{\bf p}} 
\hat{c}_{{\bf k}\uparrow}^{\dagger} \hat{c}_{{\bf p}
\downarrow}^{\dagger} \right) ,
\nonumber 
\end{eqnarray}
describing the itinerant electrons ($\hat{c}_{{\bf k}
\sigma}^{(\dagger)}$ operators) coexisting with the preformed 
pairs (bosonic $\hat{b}_{\bf q}^{(\dagger)}$ operators). They 
are mutually coupled through the charge exchange (Andreev-type) 
scattering. Such scenario (\ref{BF}) has been considered by various 
authors in the context of high $T_{c}$ superconductivity 
\cite{Ranninger-85,Friedberg-89,Micnas-90,Auerbach-02,Rice-09,
Ranninger-10,Geshkenbein-97,Senthil-09} and for description of 
the resonant Feshbach interaction in the ultracold fermion atom 
gasses \cite{Holland-01,Griffin-02,Levin-05}.

By $E_{\bf q}$ we denote the energy of preformed pairs measured 
with respect to $2\mu$. Since in the superconducting state of 
cuprate materials the energy gap $\Delta_{\bf k}$ has a $d$-wave 
symmetry we furthermore impose the anisotropic boson-fermion 
coupling $g_{{\bf k},{\bf p}}\!=\!g \left( \cos{k_x}- 
\cos{k_y}\right)$. If one restricts only to the Bose-Einstein 
condensed pairs (i.e.\ to bosonic ${\bf q}\!=\!{\bf 0}$  
mode), then the model (\ref{BF}) becomes identical with the 
reduced BCS Hamiltonian (\ref{hamil}) where $\Delta_{\bf k}$ 
is related to the condensate $-\frac{\hat{b}_{{\bf q}=
{\bf 0}}}{\sqrt{N}} \; g_{{\bf k},-{\bf k}}$. In what follows
we shall consider an influence of the non-condensed preformed 
pairs on the current-current response function.

\subsection{Outline of the continuous diagonalization}

Adopting again the Wegner's proposal \cite{Wegner-94} 
we choose the generating operator as $\hat{\eta}(l)\!=\!
[\hat{H}_0(l),\hat{V}^{B-F}(l)]$, where $\hat{H}_{0}(l)$ 
stands for the free fermion and boson contributions whereas
$\hat{V}^{B-F}(l)$ denotes their interaction term. In
explicit form, such generating operator is given by
\begin{eqnarray} 
\hat{\eta}(l)= \frac{1}{\sqrt{N}} \sum_{{\bf k},{\bf p}}
\alpha_{{\bf k},{\bf p}}(l) \left(  \hat{b}_{{\bf k}+{\bf p}}
\hat{c}_{{\bf k} \uparrow}^{\dagger} \hat{c}_{{\bf p} 
\downarrow}^{\dagger} - \mbox{h.c.}\right),
\label{eta_BF}
\end{eqnarray}
where $\alpha_{{\bf k},{\bf p}}(l)=\left( \xi_{\bf k}(l)+
\xi_{{\bf p}}(l)-E_{{\bf k}+{\bf p}}(l)\right) g_{{\bf k},
{\bf p}}(l)$. Using (\ref{eta_BF}) in the flow equation 
for the Hamiltonian (\ref{BF}) we obtain \cite{Domanski-01}
\begin{eqnarray}
\frac{d}{dl} \; \mbox{\rm ln}  g_{{\bf k},{\bf p}}(l) = 
- \left[ \xi_{\bf k}(l)\!+\!\xi_{{\bf p}}(l)
\!-\!E_{{\bf k}+{\bf p}}(l)\right]^{2} ,
\label{flow_g} 
\end{eqnarray}
which implies an exponential diminishing of 
$g_{{\bf k},{\bf p}}(l)$ and guarantees its total disappearance 
in the asymptotic limit $l\rightarrow \infty$. Simultaneously
the fermion and boson energies are renormalized 
according to the flow equations \cite{Domanski-01}
\begin{eqnarray}
\frac{d}{dl} \; \xi_{\bf k}(l) &=& 
\frac{2}{N} \sum_{\bf q} \alpha_{{\bf k},{\bf q}-{\bf k}}(l)
\; g_{{\bf k},{\bf q}-{\bf k}}(l) \; n^{B}_{\bf q}
\label{flow_xi}
\\
\frac{d}{dl} \; E_{\bf q}(l) &=& - \frac{2}{N} 
\sum_{\bf k} \alpha_{{\bf k},{\bf k}-{\bf q}}(l) \;
g_{{\bf k}-{\bf q},{\bf k}}(l) \nonumber \\ 
&& \times \left[ 1\! -\! n^{F}_{{\bf k}-{\bf q}} 
\!-\! n^{F}_{\bf k} \right]
\label{flow_E} 
\end{eqnarray}
where $n^{F/B}_{\bf k}$ denote the fermion/boson occupancies of 
momentum ${\bf k}$ state. We have previously \cite{Domanski-01,
Domanski-03} explored (analytically and numerically) the flow 
equations (\ref{flow_g}-\ref{flow_E}) 
 arriving at the following conclusions:
\begin{enumerate}
\item[{a)}] 
    the renormalized fermion dispersion $\tilde{\xi}_{\bf k}$
    develops either the true gap (below $T_{c}$,
    when a finite fraction of the BE condensed bosons exists) 
    or the pseudogap (for $T_{c}\!<\!T\!<\!T^{*}$, 
    where $T^{*}$ marks the onset of 
    superconducting-type correlations), 
\item[{b)}] 
    the long-wavelength limit of the effective boson dispersion  
    $\tilde{E}_{\bf q}$ is characterized by the Goldstone 
    mode (for $T\!<\!T_{c}$) whose remnants become overdamped 
    in the pseudogap regime (above $T_{c}$),
\item[{c)}] 
    the single particle spectral function of fermions 
    (see Appendix B) consists of the Bogolubov-type
    branches separated by the (pseudo)gap and these 
    features remain preserved up to $T^{*}$.
\end{enumerate}

More recently \cite{Ranninger-10} we have also 
investigated evolution of the ${\bf k}$-resolved pseudogap 
considering two-dimensional lattice dispersion with 
the nearest and next-nearest neighbor hopping integrals 
realistic for the cuprate superconductors. For 
temperatures slightly above $T_{c}$ we have found that 
the pseudogap starts to close around the nodal points
restoring the Fermi arcs, whereas in the antinodal 
areas the pseudogap practically does not change.
Upon a gradual increase of temperature the length 
of the Fermi arcs linearly increases, in agreement
with the experimental ARPES data \cite{Fermi_arcs}. 
Similar conclusions have been achieved for the same
model (\ref{BF}) from theoretical studies based the 
conserving diagrammatic approach \cite{Levin-09}.

\subsection{Diamagnetism due to the preexisting pairs}

Following the guidelines discussed in section II.C 
we can now formulate the linear response theory for the 
model (\ref{BF}), focusing on the role played by the 
non-condensed ${\bf q}\neq{\bf 0}$ preformed pairs.

To impose the corresponding parametrization of the current 
operators ${\bf j}^{\sigma}_{\bf q}$ we again start 
from the initial derivative (\ref{flow_j_init}). Using 
the generating operator (\ref{eta_BF}) we obtain
\begin{widetext}
\begin{eqnarray}
\left( \frac{d \; \hat{\bf j}^{\sigma}_{{\bf q}}(l)}{dl} \right)_{l=0} 
&=& \mp \sum_{\bf k} {\bf v}_{{\bf k}+\frac{\bf q}{2}} \sum_{\bf p} 
\left ( \alpha_{{\bf k},{\bf p}} \hat{b}^{\dagger}_{{\bf k}+{\bf p}}
\hat{c}_{{\bf p},-\sigma} \hat{c}_{{\bf k}+{\bf q},\sigma} 
+ \alpha_{{\bf k}+{\bf q},{\bf k}} \hat{b}_{{\bf k}+{\bf q}+{\bf p}} 
\hat{c}_{{\bf k},\sigma}^{\dagger} \hat{c}_{{\bf p},-\sigma}^{\dagger} \right ) ,
\label{j_init_BF} 
\end{eqnarray}
where $-$ ($+$) refers to the spin $\uparrow$ ($\downarrow$).
Analyzing (\ref{j_init_BF}) we deduce the following general
structure of the $l$-dependent current operators
\begin{eqnarray} 
\hat{\bf j}^{\uparrow}_{{\bf q}}(l) &=& \sum_{\bf k} 
{\bf v}_{{\bf k}+\frac{\bf q}{2}} \left( {\cal{A}}_{{\bf k},{\bf q}}(l) 
\hat{c}_{{\bf k},\uparrow}^{\dagger}\hat{c}_{{\bf k}+{\bf q},\uparrow} 
+{\cal{B}}_{{\bf k},{\bf q}}(l) \hat{c}_{-{\bf k},\downarrow} 
\hat{c}_{-({\bf k}+{\bf q}),\downarrow}^{\dagger}  \right. \nonumber \\
&& \left.
+ \sum_{\bf p}  \left( {\cal{D}}_{{\bf k},{\bf p},{\bf q}}(l)
\hat{b}_{{\bf k}+{\bf p}} \hat{c}_{{\bf k},\uparrow}^{\dagger} 
\hat{c}_{{\bf p}-{\bf q},\downarrow}^{\dagger}  
+ {\cal{F}}_{{\bf k},{\bf p},{\bf q}}(l) \hat{b}^{\dagger}_{{\bf k}+{\bf p}}
\hat{c}_{{\bf p},\downarrow} \hat{c}_{{\bf k}+{\bf q},\uparrow} 
 \right) \right)
\label{j_ansatz_BF} 
\end{eqnarray}
with the initial values ${\cal{A}}_{{\bf k},{\bf q}}(0)\!=\!1$ 
and  ${\cal{B}}_{{\bf k},{\bf q}}(0)\!=\!{\cal{D}}_{{\bf k},{\bf p},
{\bf q}}(0)\!=\!{\cal{F}}_{{\bf k},{\bf p},{\bf q}}(0)\!=\!0$.
The operator $\hat{\bf j}^{\downarrow}_{{\bf q}}(l)$ is given 
by expression analogous to (\ref{j_ansatz_BF}) with
${\cal{D}}_{{\bf k},{\bf p},{\bf q}}(l)$ replaced by 
$-{\cal{F}}_{{\bf k},{\bf p},{\bf q}}(l)$ and {\em vice versa}.
Let us remark that taking into account only the BE condensed pairs 
$\hat{b}^{(\dagger)}_{{\bf k}+{\bf p}}=\hat{b}^{(\dagger)}_{\bf 0}
\delta_{{\bf p},-{\bf k}}$ we would come back to the ansatz 
(\ref{j_ansatz}) reproducing the BCS solution.

After a somewhat lengthy but rather straightforward algebra we 
derive the following set of the flow equations
\begin{eqnarray}
\frac{d {\cal{A}}_{{\bf k},{\bf q}}(l)}{dl} &=&
\sum_{\bf p} \left[ \alpha_{{\bf k}+{\bf q},{\bf p}-{\bf q}}(l)
{\cal{D}}_{{\bf k},{\bf p},{\bf q}}(l) \left( n^{F}_{{\bf p}-{\bf q}}
+ n^{B}_{{\bf k}+{\bf p}} \right) + \alpha_{{\bf k},{\bf p}}(l)
{\cal{F}}_{{\bf k},{\bf p},{\bf q}}(l) \left( n^{F}_{{\bf p}}
+ n^{B}_{{\bf k}+{\bf p}} \right) \right]
\label{flow_A} \\
\frac{d {\cal{B}}_{{\bf k},{\bf q}}(l)}{dl} &=&
- \sum_{\bf p} \left[ \alpha_{{\bf k},{\bf p}}(l)
{\cal{D}}_{-{\bf p},-{\bf k},{\bf q}}(l) \left( n^{F}_{{\bf p}}
+ n^{B}_{{\bf k}+{\bf p}} \right) + \alpha_{{\bf k}+{\bf q},
{\bf p}-{\bf q}}(l) {\cal{F}}_{-{\bf p},-{\bf k},{\bf q}}(l) 
\left( n^{F}_{{\bf p}-{\bf q}}+ n^{B}_{{\bf k}+{\bf p}} 
\right) \right]
\label{flow_B} \\
\frac{d {\cal{D}}_{{\bf k},{\bf p},{\bf q}}(l)}{dl} &=&
- \alpha_{{\bf k}+{\bf q},{\bf p}-{\bf q}}(l)
{\cal{A}}_{{\bf k},{\bf q}}(l)
+ \alpha_{{\bf k},{\bf p}}(l)
{\cal{B}}_{-{\bf p},{\bf q}}(l) ,
\label{flow_D} \\
\frac{d {\cal{F}}_{{\bf k},{\bf p},{\bf q}}(l)}{dl} &=&
- \alpha_{{\bf k},{\bf p}}(l)
{\cal{A}}_{{\bf k},{\bf q}}(l)
+ \alpha_{{\bf k}+{\bf q},{\bf p}-{\bf q}}(l)
{\cal{B}}_{-{\bf p},{\bf q}}(l) .
\label{flow_F}
\end{eqnarray}
For deriving the equations (\ref{flow_A},\ref{flow_B}) we used 
the following approximations
\begin{eqnarray}
\hat{b}^{\dagger}_{\bf k} \hat{b}_{{\bf k}'} \;
\hat{c}^{\dagger}_{{\bf p},\sigma} \hat{c}_{{\bf p}',{\sigma}}
& \simeq & \delta_{{\bf k},{\bf k}'} n^{B}_{\bf k} \;
\hat{c}^{\dagger}_{{\bf p},\sigma} \hat{c}_{{\bf p}',{\sigma}}
+\hat{b}^{\dagger}_{\bf k} \hat{b}_{{\bf k}'} \;
\delta_{{\bf p},{\bf p}'} n^{F}_{\bf p} 
- \delta_{{\bf k},{\bf k}'} n^{B}_{\bf k} \;
\delta_{{\bf p},{\bf p}'}n^{F}_{\bf p}
\label{eqn42}
\\
\hat{c}^{\dagger}_{{\bf p},\uparrow} \hat{c}_{{\bf p}',{\uparrow}} \;
\hat{c}^{\dagger}_{{\bf p},\downarrow} \hat{c}_{{\bf p}',{\downarrow}}
& \simeq & \delta_{{\bf k},{\bf k}'} n^{F}_{\bf k} \;
\hat{c}^{\dagger}_{{\bf p},\downarrow} \hat{c}_{{\bf p}',{\downarrow}}
+\hat{c}^{\dagger}_{{\bf p},\uparrow} \hat{c}_{{\bf p}',{\uparrow}} \;
\delta_{{\bf p},{\bf p}'} n^{F}_{\bf p} 
- \delta_{{\bf k},{\bf k}'} n^{F}_{\bf k} \;
\delta_{{\bf p},{\bf p}'}n^{F}_{\bf p}
\label{eqn43}
\end{eqnarray}
neglecting the higher order products $\delta \hat{X} \; \delta \hat{Y}$ 
of the fluctuations $\delta \hat{X}= \hat{X} -\langle \hat{X} \rangle$, 
where the corresponding observables for the case of Eq.\ (\ref{eqn42}) 
are defined by 
$\hat{X}=\hat{b}^{\dagger}_{\bf k} \hat{b}_{{\bf k}'}$, $\hat{Y}=
\hat{c}^{\dagger}_{{\bf p},\sigma} \hat{c}_{{\bf p}',{\sigma}}$ 
and for (\ref{eqn43})  by
$\hat{X}=\hat{c}^{\dagger}_{{\bf p},\uparrow} \hat{c}_{{\bf p}'
,{\uparrow}}$, $\hat{Y}=\hat{c}^{\dagger}_{{\bf p},\downarrow} 
\hat{c}_{{\bf p}',{\downarrow}}$. Such truncations (\ref{eqn42},
\ref{eqn43}) enable us to satisfy the flow equation $\frac{d}{dl} 
\; \hat{\bf j}^{\sigma}_{{\bf q}}(l)=\left[ \hat{\eta}(l),
\hat{\bf j}^{\sigma}_{{\bf q}}(l)\right]$ using the 
parametrization (\ref{j_ansatz_BF}) imposed on the current 
operators $\hat{\bf j}^{\sigma}_{{\bf q}}(l)$. Otherwise, if  
we introduced these neglected terms $\hat{X} \hat{Y}$ to the $l$-dependent 
operator $\hat{\bf j}^{\sigma}_{{\bf q}}(l)$ they would induce 
even more complex structures arising from the commutator $\left[ 
\hat{\eta}(l),\hat{\bf j}^{\sigma}_{{\bf q}}(l)\right]$ and
formally the flow equation (\ref{O_flow}) could never be
 obeyed (except for only the exactly solvable cases).
The truncations (\ref{eqn42},\ref{eqn43}) or similar, 
represent thus a necessary compromise in which the flow 
equation technique deals with the physical problems which 
are not exactly solvable \cite{Wegner-94,Wilson-94}.

Finally, let us determine the current-current response function 
(\ref{jj}), keeping in mind that the statistical averaging is 
feasible with respect to $\hat{H}(\infty)$. Using 
the ansatz (\ref{j_ansatz_BF}) we find the response function  
\begin{eqnarray}
\langle \langle \hat{j}_{{\bf q},\alpha}; \hat{j}_{-{\bf q},\beta} 
\rangle\rangle &=& \sum_{{\bf k},{\bf p}} v_{{\bf k}+\frac{\bf q}{2},
\alpha} v_{{\bf p}-\frac{\bf q}{2},\beta} \left( \tilde{\cal{A}}_{{\bf k},
{\bf q}} \tilde{\cal{A}}_{{\bf p},-{\bf q}} 
\sum_{\sigma} \langle \langle \hat{c}^{\dagger}_{{\bf k},\sigma} 
\hat{c}_{{\bf k}+{\bf q},\sigma} ;\hat{c}^{\dagger}_{{\bf p},
\sigma} \hat{c}_{{\bf p}-{\bf q},\sigma} \rangle\rangle 
\right.  \nonumber \\ && + \; 
\tilde{\cal{A}}_{{\bf k},{\bf q}} \tilde{\cal{B}}_{{\bf p},-{\bf q}}
 \sum_{\sigma} \langle \langle \hat{c}^{\dagger}_{{\bf k},\sigma} 
\hat{c}_{{\bf k}+{\bf q},\sigma} ; \hat{c}_{-{\bf p},\sigma} 
\hat{c}^{\dagger}_{-({\bf p}-{\bf q}),\sigma} \rangle\rangle 
\nonumber \\ && + \; 
\tilde{\cal{B}}_{{\bf k},{\bf q}} \tilde{\cal{A}}_{{\bf p},-{\bf q}}
 \sum_{\sigma} \langle \langle \hat{c}_{-{\bf k},\sigma} 
\hat{c}^{\dagger}_{-({\bf k}+{\bf q}),\sigma} ;\hat{c}^{\dagger}
_{{\bf p},\sigma} \hat{c}_{{\bf p}-{\bf q},\sigma} \rangle\rangle 
\nonumber \\ && + \;
\tilde{\cal{B}}_{{\bf k},{\bf q}} \tilde{\cal{B}}_{{\bf p},-{\bf q}}
 \sum_{\sigma} \langle \langle \hat{c}_{-{\bf k},\sigma} 
\hat{c}^{\dagger}_{-({\bf k}+{\bf q}),\sigma} ; \hat{c}_{-{\bf p},
\sigma} \hat{c}^{\dagger}_{-({\bf p}-{\bf q}),\sigma} \rangle
\rangle \nonumber \\ && - \; 
\sum_{{\bf k}',{\bf p}'} \tilde{\cal{G}}_{{\bf k},{\bf k}',{\bf q}}
 \tilde{\cal{G}}_{{\bf p},{\bf p}',-{\bf q}} \langle 
\langle \hat{b}_{{\bf k}+{\bf k}'}\hat{c}^{\dagger}_{{\bf k},
\uparrow} \hat{c}^{\dagger}_{{\bf k}'-{\bf q},\downarrow} ;
\hat{b}^{\dagger}_{{\bf p}+{\bf p}'} \hat{c}_{{\bf p}',\downarrow} 
\hat{c}_{{\bf p}-{\bf q},\uparrow} \rangle\rangle 
\nonumber \\ && - \; 
\sum_{{\bf k}',{\bf p}'} \tilde{\cal{G}}_{{\bf p},{\bf p}',{\bf q}}
\tilde{\cal{G}}_{{\bf k},{\bf k}',-{\bf q}} \langle 
\langle \hat{b}^{\dagger}_{{\bf p}+{\bf p}'} \hat{c}_{{\bf p}',
\downarrow} \hat{c}_{{\bf p}+{\bf q},\uparrow} ; 
\hat{b}^{\dagger}_{{\bf p}+{\bf p}'} \hat{c}_{{\bf p}',
\downarrow} \hat{c}_{{\bf p}+{\bf q},\uparrow}
\rangle\rangle  
\label{6_contributions}
\end{eqnarray}
where $\tilde{\cal{G}}_{{\bf p},{\bf p}',{\bf q}} \equiv
\tilde{\cal{D}}_{{\bf p},{\bf p}',{\bf q}} - \tilde{\cal{F}}
_{{\bf p},{\bf p}',{\bf q}}$ and we used the abbreviation
$\langle \langle \hat{O}_{1};\hat{O}_{2} \rangle\rangle 
\equiv - \langle \hat{T}_{\tau}  \hat{O}_{1}(\tau)
\hat{O}_{2} \rangle_{\hat{H}(\infty)}$.
These contributions (\ref{6_contributions}) are depicted 
graphically in figure \ref{plot1}. Vertices denoted by 
the filled squares correspond to the asymptotic 
value $\tilde{\cal{G}}$ whereas the filled circles 
represent $\tilde{\cal{A}}$ and/or $\tilde{\cal{B}}$.

\begin{figure}
\hspace{-0.5cm}
\epsfxsize=5.1cm{\epsffile{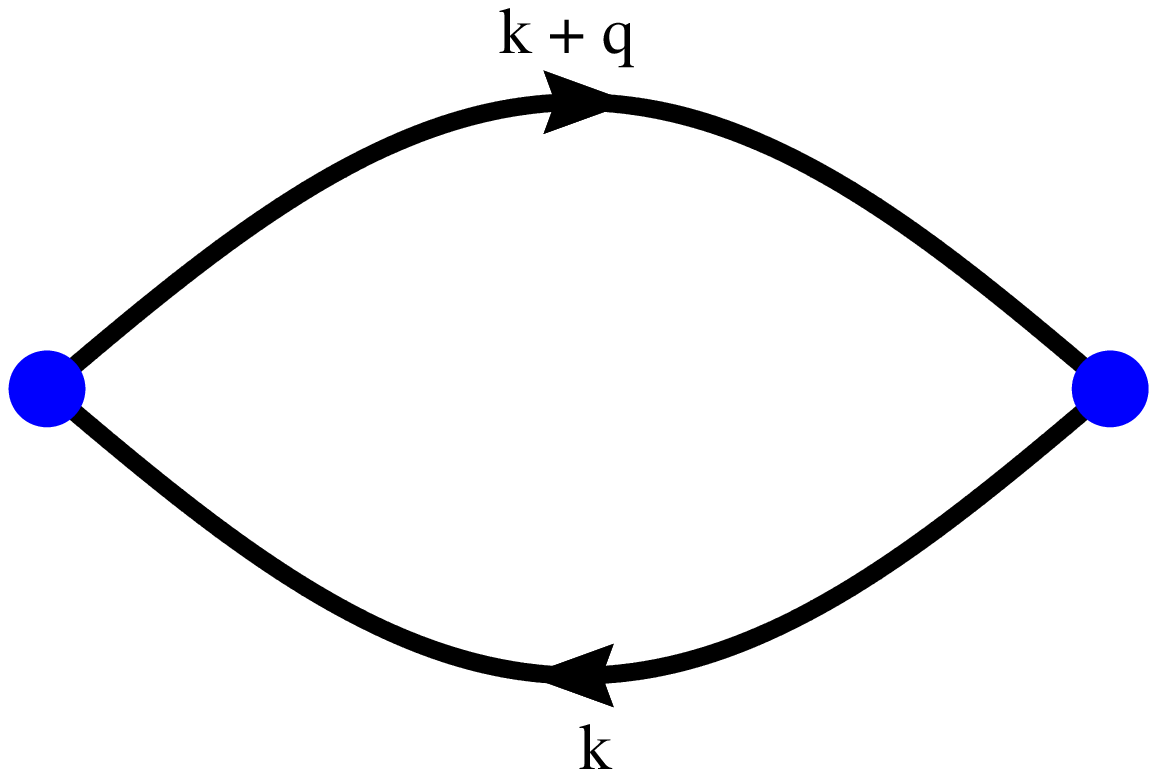}}
\hspace{-0.1cm}
\epsfxsize=5.5cm{\epsffile{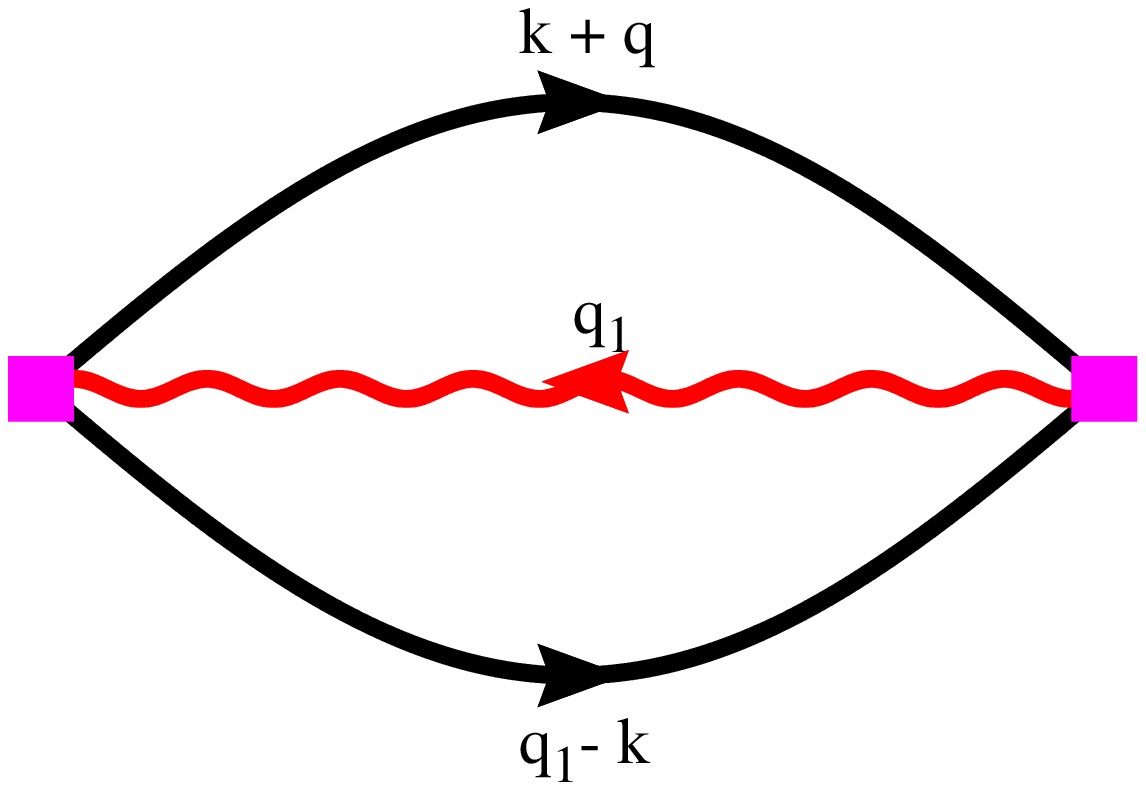}}
\hspace{-0.1cm}
\epsfxsize=5.5cm{\epsffile{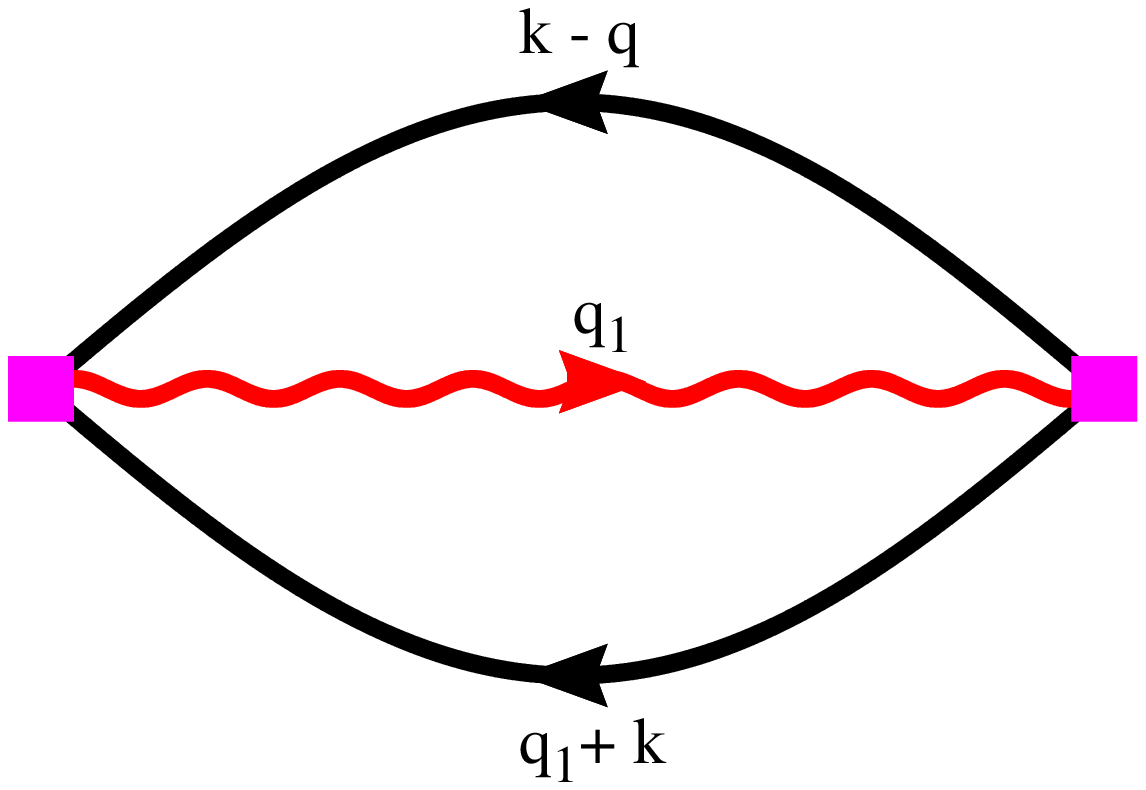}}
\caption{(color online) Contributions to the current-current 
response function from the particle-hole bubble (the left h.s.\ 
panel) and from additional terms involving the finite momentum 
boson propagators (the middle and right h.s.\ panels). Vertices
are expressed by the corresponding momentum components of the 
asymptotic values $l \rightarrow \infty$ for the coefficients 
used in Eq.\ (\ref{j_ansatz_BF}).}
\label{plot1}
\end{figure}

Performing the Matsubara summation for the particle-hole 
convolutions (left panel in figure \ref{plot1}) and the 
double Matsubara summation for the diagrams involving one 
bosonic and two fermionic propagators we obtain the following
Fourier transform of (\ref{6_contributions}) 
\begin{eqnarray}
\Pi_{\alpha,\beta}({\bf q},i\nu) &=& \sum_{\bf k} v_{{\bf k}
+\frac{\bf q}{2},\alpha} v_{{\bf k}+\frac{\bf q}{2},\beta}   
\left\{ \left[ \tilde{\cal{A}}_{{\bf k},{\bf q}}  
\tilde{\cal{A}}_{{\bf k}+{\bf q},-{\bf q}} 
+\tilde{\cal{A}}_{{\bf k},{\bf q}}  
\tilde{\cal{B}}_{-{\bf k},-{\bf q}}
\right. \right. \label{response_BF} \\ &+& \left. \left.
\tilde{\cal{A}}_{-{\bf k},-{\bf q}}  
\tilde{\cal{B}}_{{\bf k},{\bf q}}
+\tilde{\cal{B}}_{{\bf k},{\bf q}}  
\tilde{\cal{B}}_{{\bf k}+{\bf q},-{\bf q}}\right]
\left[ f_{FD}(\tilde{\xi}_{{\bf k}+{\bf q}}) \!-\! 
f_{FD}(\tilde{\xi}_{\bf k}) \right]
\left[ \frac{1} {i \nu \!+\! \tilde{\xi}_{\bf k+q} 
\!-\! \tilde{\xi}_{\bf k}} - \frac{1} {i \nu\! -\! 
\tilde{\xi}_{\bf k+q}\! +\! \tilde{\xi}_{\bf k}} \right] 
\right. \nonumber \\ &+& \left.
 \sum_{{\bf k}'} \tilde{\cal{G}}_{{\bf k},-{\bf k}',{\bf q}} 
\tilde{\cal{G}}_{{\bf k}+{\bf q},-({\bf k}'+{\bf q}),-{\bf q}}
\left(  \left[ 1 \!-\! f_{FD}(\tilde{\xi}_{{\bf k}+{\bf q}}) 
\!-\! f_{FD}(\tilde{\xi}_{{\bf k}'}) \right] 
\frac{f_{BE}(\tilde{E}_{{\bf k}-{\bf k}'})-f_{BE}(\tilde
{\xi}_{{\bf k}+{\bf q}}+\tilde{\xi}_{{\bf k}'})}
{i \nu \!-\! (\tilde{\xi}_{\bf k+q}\!+\! \tilde{\xi}_{{\bf k}'}-
\tilde{E}_{{\bf k}-{\bf k}'})} 
\right. \right. \nonumber  \\ && 
\hspace{4cm} -\left. \left. 
 \left[ 1 \!-\! f_{FD}(\tilde{\xi}_{{\bf k}'+{\bf q}}) 
\!-\! f_{FD}(\tilde{\xi}_{\bf k}) \right]
\frac{f_{BE}(\tilde{E}_{{\bf k}-{\bf k}'})-f_{BE}(\tilde
{\xi}_{{\bf k}'+{\bf q}}+\tilde{\xi}_{{\bf k}})}
{i \nu \!+\! (\tilde{\xi}_{{\bf k}'+{\bf q}}\!+\! 
\tilde{\xi}_{{\bf k}}-\tilde{E}_{{\bf k}-{\bf k}'})} 
 \right) \right\} .  
\nonumber
\end{eqnarray}
where $f_{BE}(\omega)\!=\!\left[\mbox{\rm exp}(\omega/k_{B}T)
\!-\!1\right]^{-1}$ is the Bose-Einstein distribution. The 
function (\ref{response_BF}) in a straightforward manner 
generalizes the previous BCS form (\ref{BCS_result}) and 
is the central result of our study.

The {\em d.c.} diamagnetic properties of the system 
depend on the static value of the response function.
In our present case it is given by
\begin{eqnarray}
\Pi_{\alpha,\alpha}({\bf q},i\nu\!=\!0) &=& 
\sum_{\bf k} v^{2}_{{\bf k} + \frac{\bf q}{2},\alpha} 
\left\{ 2 \left[ \tilde{\cal{A}}_{{\bf k},{\bf q}}  
\tilde{\cal{A}}_{{\bf k}+{\bf q},-{\bf q}} 
+\tilde{\cal{A}}_{{\bf k},{\bf q}}  
\tilde{\cal{B}}_{-{\bf k},-{\bf q}}
+ \tilde{\cal{A}}_{-{\bf k},-{\bf q}}  
\tilde{\cal{B}}_{{\bf k},{\bf q}}
+\tilde{\cal{B}}_{{\bf k},{\bf q}}  
\tilde{\cal{B}}_{{\bf k}+{\bf q},-{\bf q}}\right]
\frac{ f_{FD}(\tilde{\xi}_{{\bf k}+{\bf q}}) \!-\! 
f_{FD}(\tilde{\xi}_{\bf k}) }{
\tilde{\xi}_{\bf k+q} \!-\! \tilde{\xi}_{\bf k}} 
\right. \nonumber \\ &+&  
 \sum_{{\bf k}'} \tilde{\cal{G}}_{{\bf k},-{\bf k}',{\bf q}} 
\tilde{\cal{G}}_{{\bf k}+{\bf q},-({\bf k}'+{\bf q}),-{\bf q}}
\left(  \left[ f_{BE}(\tilde{E}_{{\bf k}-{\bf k}'})
-f_{BE}(\tilde{\xi}_{{\bf k}+{\bf q}}+\tilde{\xi}_{{\bf k}'}) 
\right] \frac{1 \!-\! f_{FD}(\tilde{\xi}_{{\bf k}+{\bf q}}) 
\!-\! f_{FD}(\tilde{\xi}_{\bf k}')}
{\tilde{E}_{{\bf k}-{\bf k}'}-(\tilde{\xi}_{\bf k+q}
\!+\! \tilde{\xi}_{{\bf k}'})} \right.  \nonumber  \\ && 
\hspace{4cm} + \left. \left. 
 \left[ f_{BE}(\tilde{E}_{{\bf k}-{\bf k}'})-f_{BE}(\tilde
{\xi}_{{\bf k}'+{\bf q}}+\tilde{\xi}_{{\bf k}}) \right]
\frac{1 \!-\! f_{FD}(\tilde{\xi}_{{\bf k}'+{\bf q}}) 
\!-\! f_{FD}(\tilde{\xi}_{\bf k})}
{\tilde{E}_{{\bf k}-{\bf k}'}-(\tilde{\xi}_{{\bf k}'+{\bf q}}
\!+\! \tilde{\xi}_{{\bf k}})} 
 \right) \right\}.  
\label{static_limit}
\end{eqnarray}
\end{widetext}
For temperatures below $T_{c}$ (when a finite fraction 
of the Bose-Einstein condensed pairs exists) the main 
contribution in the expression  (\ref{static_limit}) 
comes from ${\bf k}'\!=\!{\bf k}$ terms. Under such 
conditions Eq.\ (\ref{static_limit}) becomes identical
with the BCS solution consisting of: 
a) the {\em superfluid} fraction [i.e.\ the term  in Eq.\ 
(\ref{BCS_result}) proportional to the coherence factor 
$(\tilde{u}_{\bf k+q} \tilde{v}_{\bf k} - \tilde{v}_{
\bf k+q} \tilde{u}_{\bf k})^{2}$], and the other 
b) {\em normal} contribution from the thermally excited 
quasiparticles, i.e.\ the term proportional to 
$(\tilde{u}_{\bf k+q} \tilde{u}_{\bf k} + 
\tilde{v}_{\bf k+q} \tilde{v}_{\bf k})^{2}$  
\cite{Schrieffer_book}.

Above $T_{c}$ (but fairly below $T^{*}$) a considerable 
amount of the preformed pairs occupies the low momenta 
states $E_{{\bf q} \sim {\bf 0}}$ therefore expression 
(\ref{static_limit}) becomes reminiscent of the above 
mentioned BCS components in the response function.

\section{Iterative solution}

To explore the physical aspects related to the current-current response 
function (\ref{response_BF}) we adopt an iterative method for solving 
the coupled flow equations (\ref{flow_A}-\ref{flow_F}). Such scheme 
allows for an approximate  estimation of the introduced $l$-dependent
parameters. Following \cite{Domanski-01} we make use of the fact 
that the dominant renormalization affects the boson-fermion coupling 
$g_{{\bf k},{\bf p}}(l)$ which ultimately vanishes in the asymptotic 
limit $l\rightarrow\infty$. Neglecting the simultaneous renormalization 
of the fermion $\xi_{\bf k}(l) \simeq \xi_{\bf k}$ and boson 
energies $E_{\bf q}(l) \simeq E_{\bf q}$ we obtain the following 
solution of the flow equation (\ref{flow_g})
\begin{eqnarray}
 g_{{\bf k},{\bf p}}(l) \simeq  g_{{\bf k},{\bf p}}
e^{- \left( \xi_{\bf k}+\xi_{\bf p}-E_{{\bf k}+{\bf p}}
\right)^{2} l} .
\label{appr_g} 
\end{eqnarray}
Substituting this result (\ref{appr_g}) to the flow equations 
(\ref{flow_xi},\ref{flow_E}) we might in turn update the energies and 
the routine can be continued, at each  iterative level providing 
a better and better estimation for the renormalized 
quantities.

In this section we apply such scheme to the flow equations 
(\ref{flow_A}-\ref{flow_F}), restricting ourselves to the
lowest order solutions based on Eq.\ (\ref{appr_g}). We start 
with the initial values of the coefficients ${\cal{A}}_{{\bf k},
{\bf q}}(l) \simeq {\cal{A}}_{{\bf k},{\bf q}}(0)\!=\!1$ and
${\cal{B}}_{{\bf k},{\bf q}}(l) \simeq {\cal{B}}_{{\bf k},
{\bf q}}(0)\!=\!0$ substituting them in the right h.s.\ of 
the flow equations (\ref{flow_D},\ref{flow_F}). Using Eq.\ 
(\ref{appr_g}) we analytically solve the simplified equations
(\ref{flow_D},\ref{flow_F}) obtaining
\begin{eqnarray}
{\cal{D}}_{{\bf k},{\bf p},{\bf q}}(l) &\simeq&
\frac{g_{{\bf k}+{\bf q},{\bf p}-{\bf q}} \left[
e^{-\left( \xi_{{\bf k}\!+\!{\bf q}}\!+\!\xi_{{\bf p}
\!-\!{\bf q}}\!-\!E_{{\bf k}\!+\!{\bf p}} \right)^{2}l}
\!-\!1 \right]}{\xi_{{\bf k}+{\bf q}}
+\xi_{{\bf p}-{\bf q}}-E_{{\bf k}+{\bf p}}} ,
\label{appr_D} \\
{\cal{F}}_{{\bf k},{\bf p},{\bf q}}(l) &\simeq&
\frac{g_{{\bf k},{\bf p}} \left[ e^{-\left( \xi_{{\bf k}}
+\xi_{{\bf p}}-E_{{\bf k}+{\bf p}} \right)^{2}l} -1 
\right]}{\xi_{{\bf k}}+\xi_{{\bf p}}-E_{{\bf k}+{\bf p}} } .
\label{appr_F} 
\end{eqnarray}
Their asymptotic values are  given by
\begin{eqnarray}
\tilde{\cal{D}}_{{\bf k},{\bf p},{\bf q}} &\simeq&
- \; \frac{g_{{\bf k}+{\bf q},{\bf p}-{\bf q}}}{\xi_{{\bf k}
+{\bf q}}+\xi_{{\bf p}-{\bf q}}-E_{{\bf k}+{\bf p}} } ,
\label{appr_D_inf} \\
\tilde{\cal{F}}_{{\bf k},{\bf p},{\bf q}} &\simeq& - \;
\frac{g_{{\bf k},{\bf p}} }{\xi_{{\bf k}}
+\xi_{{\bf p}}-E_{{\bf k}+{\bf p}} } .
\label{appr_F_inf} 
\end{eqnarray}
Using the $l$-dependent coefficients ${\cal{D}}_{{\bf k},
{\bf p},{\bf q}}(l)$ and ${\cal{F}}_{{\bf k},{\bf p},
{\bf q}}(l)$ we can next determine ${\cal{A}}_{{\bf k},
{\bf p}}(l)$ and ${\cal{B}}_{{\bf k},{\bf p}}(l)$. 
Substituting (\ref{appr_D},\ref{appr_F}) in the right 
h.s.\ of (\ref{flow_A},\ref{flow_B}) we obtain the 
following asymptotic values
\begin{eqnarray}
\tilde{\cal{A}}_{{\bf k},{\bf q}} &\simeq& 1 -
\frac{1}{2} \sum_{\bf p} \left[ 
\frac{ \left( n^{F}_{{\bf p}}+ n^{B}_{{\bf k}+{\bf p}} 
 \right)|g_{{\bf k},{\bf p}}|^{2}}
{\left( \xi_{{\bf k}} + \xi_{{\bf p}} 
- E_{{\bf k}+{\bf p}} \right)^{2}}
\right. \nonumber \\ &+& \left.
\frac{ \left(
 n^{F}_{{\bf p}-{\bf q}}+ n^{B}_{{\bf k}+{\bf p}} 
 \right)|g_{{\bf k}+{\bf q},{\bf p}-{\bf q}}|^{2}}
{\left( \xi_{{\bf k}+{\bf q}} + \xi_{{\bf p}
-{\bf q}} - E_{{\bf k}+{\bf p}} \right)^{2}}
\right]
\label{appr_A_inf} 
\end{eqnarray}
and
\begin{eqnarray}
&&\tilde{\cal{B}}_{{\bf k},{\bf q}} \simeq
\sum_{\bf p}  g_{{\bf k},{\bf p}} g_{{\bf k}+
{\bf q},{\bf p}-{\bf q}}   \times
\label{appr_B_inf}  
\\ &&  \left[ \frac{ n^{F}_{{\bf p}} 
+ n^{B}_{{\bf k}+{\bf p}}}{
X_{{\bf k}+{\bf q},{\bf p}-{\bf q}}}
\left( \frac{1}{X_{{\bf k},{\bf p}}} - 
\frac{X_{{\bf k},{\bf p}}}{X^{2}_{{\bf k},
{\bf p}}+X^{2}_{{\bf k}+{\bf q},{\bf p}-{\bf q}}}
\right) +\right. \nonumber \\ && 
\left. \frac{ n^{F}_{{\bf p}
-{\bf q}} + n^{B}_{{\bf k}+{\bf p}} }{X_{{\bf k},{\bf p}}}
\left( \frac{1}{X_{{\bf k}+{\bf q},{\bf p}-{\bf q}}} 
-  \frac{X_{{\bf k}+{\bf q},{\bf p}-{\bf q}}}{X^{2}_{{\bf k},{\bf p}}+
X^{2}_{{\bf k}+{\bf q},{\bf p}-{\bf q}}}
\right) \right]  \nonumber
\end{eqnarray}
where $X_{{\bf k},{\bf p}}\!\equiv\!
\xi_{\bf k}\!+\!\xi_{\bf p}\!-\!E_{{\bf k}+{\bf p}}$.

Since eventual diamagnetism is determined by the long wavelength 
limit of the static response function (\ref{static_limit}) we 
focus on ${\bf q}\!=\!{\bf 0}$ values of the coefficients. 
Examining the ${\bf q}\!\rightarrow\!{\bf 0}$ limit of 
the asymptotic values (\ref{appr_D_inf},\ref{appr_F_inf}) 
we notice that the {\em superfluid} vertices vanish 
\begin{eqnarray}
\tilde{G}_{{\bf k},{\bf p},{\bf q}}\!=\!
\tilde{D}_{{\bf k},{\bf p},{\bf q}}
\!-\!\tilde{F}_{{\bf k},{\bf p},{\bf q}}
\stackrel{{\bf q} = {\bf 0}}{\longrightarrow} 0 
\end{eqnarray}
and (similarly to the BCS treatment \cite{Schrieffer_book})
we are left only with the {\em normal} component of
the paramagnetic term 
\begin{eqnarray}
\lim_{{\bf q} \rightarrow {\bf 0}} 
\Pi_{\alpha,\alpha}({\bf q},0) \!\simeq\! 
2 \sum_{\bf k} v^{2}_{{\bf k},\alpha} 
\left[ {\cal{A}}_{{\bf k},{\bf 0}} \!+\! 
{\cal{B}}_{{\bf k},{\bf 0}} \right]^{2}
\frac{d f_{FD}(\tilde{\xi}_{\bf k}) }{d \tilde{\xi}_{\bf k}} .
 \label{dia_resp}
\end{eqnarray}

At high temperatures (in a normal state) the dispersion 
$\tilde{\xi}_{\bf k}$  is nearly identical with the bare 
energy $\varepsilon_{\bf k}\!-\!\mu$ therefore Eq.\ 
(\ref{dia_resp}) cancels out the diamagnetic term of 
the response kernel $K_{\alpha,\beta}({\bf q}\rightarrow
{\bf 0},0)$ and consequently the system does not show any 
diamagnetic features. On the other hand, in the superconducting 
state the single particle excitations $\tilde{\xi}_{\bf k}$ 
are gaped and at low temperatures  $\frac{d}{d \tilde{\xi}
_{\bf k}}f_{FD}(\tilde{\xi}_{\bf k}) \approx -\delta 
(\tilde{\xi}_{\bf k})$ therefore the  paramagnetic 
contribution (\ref{dia_resp}) vanishes \cite{Schrieffer_book}. 
One then obtains a perfect diamagnetism with the characteristic 
London penetration depth  $\lambda_{L}^{-2}=ne^{2}/m$. 
Between these extreme regimes we can expect an intermediate behavior. 
In particular, for temperatures $T_{c}<T<T^{*}$ the single particle 
fermion spectrum becomes partly depleted around the Fermi energy 
so the paramagnetic term (\ref{dia_resp}) would no longer be 
able to compensate completely the diamagnetic contribution 
generating a fragile diamagnetism.

\begin{figure}
\epsfxsize=10cm{\epsffile{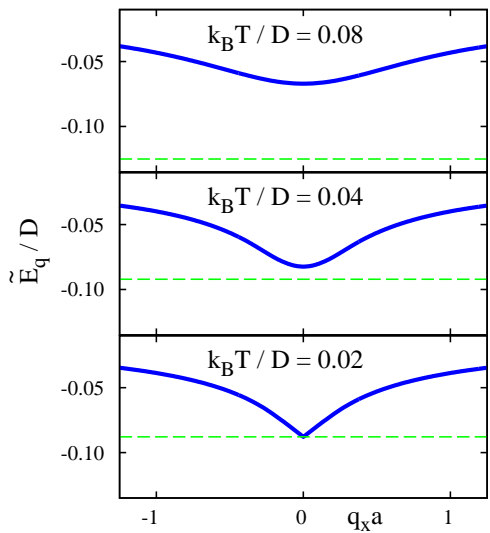}}
\caption{(color online) The renormalized boson energy 
$\tilde{E}_{\bf q}$ obtained for three representative 
temperature regions: $T>T^{*}$ (top panel), $T^{*}>T>T_{c}$ 
(middle plot) and $T_{c}>T$ (bottom panel). The dashed 
lines show the level $2\mu(T)$.}
\label{boson_changes}
\end{figure}

For some quantitative illustration of this behavior 
we have analyzed temperature dependence of the superfluid 
fraction $n_{s}(T)$ defined by the relation \cite{Fetter_book}
\begin{eqnarray}
{J}_{x}({\bf q}\rightarrow{\bf 0},0) = -\; 
\frac{e^2n_{s}(T)}{m} {A}_{x}({\bf q}\rightarrow{\bf 0},0)   .
\label{superfluid}
\end{eqnarray} 
In order to
determine $n_{s}(T)$ we substituted the paramagnetic term 
(\ref{dia_resp}) to the kernel function $K_{\alpha,\beta}
({\bf q}\rightarrow{\bf 0},0)$ and applied the coefficients 
(\ref{appr_A_inf},\ref{appr_B_inf}), simplifying the fermion 
and boson concentrations by $n^{F}_{{\bf k}}\approx  
f_{FD}(\tilde{\xi}_{\bf k})$  and $n^{B}_{{\bf q}}\approx  
f_{BE}(\tilde{E}_{\bf q})$. Furthermore, we replaced 
all energies by the renormalized values $\tilde{\xi}_{\bf k}$  
and $\tilde{E}_{\bf q}$ to account for the iterative feedback 
effects. Following the previous study \cite{Ranninger-10} we 
have selfconsistently determined these renormalized energies 
$\tilde{\xi}_{\bf k}$, $\tilde{E}_{\bf q}$ by solving 
numerically the flow equations (\ref{flow_xi},\ref{flow_E}) 
for the tight-binding lattice model $\varepsilon_{\bf k}=
-2t \left[ \cos{ak_x}+\cos{ak_y}\right] - 2t_{z} \cos{ck_z}$ 
assuming reduced mobility along $z$-axis $t_z=0.1t$. 
Initially (at $l\!=\!0$) we have assumed bosons to be 
localized. To establish some correspondence with the recent 
QMC studies \cite{Troyer-11} we have used the same total 
concentration of carriers $0.16$ and imposed the coupling 
$g=0.2D$ (where $D=8t$).

\begin{figure}
\epsfxsize=8.5cm{\epsffile{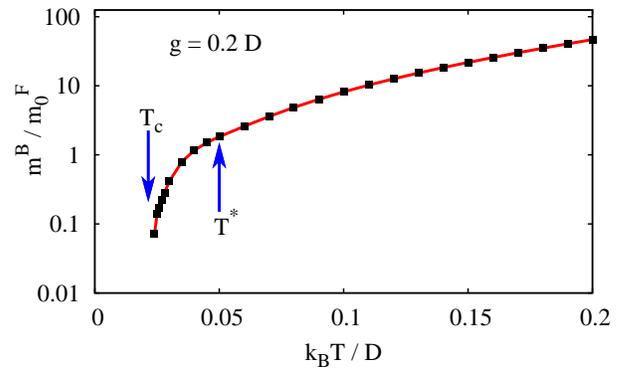}}
\caption{(color online) Temperature dependence of the effective 
boson mass $m^{B}$ obtained for the initially discrete energy 
level $E_{\bf q}(l\!=\!0)\!=\!const$. Our results resemble 
the QMC data shown in Fig.\ 9a of Ref.\ \cite{Troyer-11}.}
\label{boson_mass}
\end{figure}

The important changeover of the boson dispersion $\tilde{E}_{\bf q}$ 
upon varying temperature is shown in figure \ref{boson_changes}. 
We noticed that below some characteristic temperature $k_{B}T^{*} 
\sim 0.05D$ there occurred a considerable reduction of the in-plane 
boson mass, defined as $d^{2} \tilde{E}_{\bf q} /dq_{x}^{2} = 
\hbar^{2}/m^{B}$. Its temperature dependence [compared to the 
bare planar mass of fermions $m^{F}_{0}=\hbar^{2}/2ta^{2}$] is 
illustrated in figure \ref{boson_mass}. The mentioned suppression 
of the boson mass below $T^{*}$ coincided with appearance of 
the pseudogap in the fermion spectrum near $\mu$ -- this property
has been discussed at length in our previous studies 
\cite{Domanski-01,Domanski-03} where we formulated the flow 
equation procedure for the present model (\ref{BF}). 
Below the other temperature  $k_{B}T_{c} \sim 0.026 D$ 
the Bose-Einstein (BE) condensate appeared in the system and 
simultaneously the parabolic dispersion evolved into the 
collective sound-wave mode $\tilde{E}_{\bf q} \propto 
|{\bf q}|$ (see the bottom panel in Fig.\ \ref{boson_changes}). 

\begin{figure}
\epsfxsize=8.5cm{\epsffile{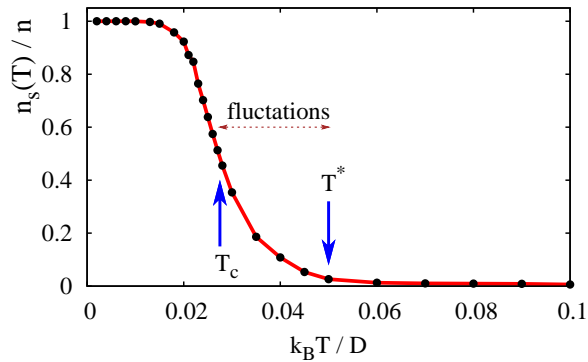}}
\caption{(color online) Temperature dependence of the superfluid
fraction $n_{s}(T)$ normalized to the total fermion concentration
$n$. We can notice a broad regime where the fragile Meissner 
effect is caused by the superconducting fluctuations.}
\label{sc_fraction}
\end{figure}

Evolution of the effective boson and fermion spectra revealed
a substantial influence on the superfluid fraction. In figure 
\ref{sc_fraction} we show the temperature dependence of such $n_{s}
(T)$. Below the temperature $T^{*}$ (for here chosen set of 
the model parameters $T^{*} \sim 2T_{c}$) we observed a gradual 
buildup of the superfluid fraction. Passing below $T_{c}$ the 
superfluid fraction exhibited a further, stronger enhancement 
manifesting an onset of the long-range phase coherence 
caused by appearance of the Bose Einstein condensate of pairs. 
At still lower temperatures, i.e.\ deep in the superconducting 
state $T \ll T_{c}$, we observed some flattening of 
the superfluid density, rather than the expected linear dependence
$n_{s}(T \rightarrow 0) \simeq n_{s}(0)-\alpha T$  
typical for $d$-wave superconducting systems with the Dirac-like 
excitations around the nodal points \cite{Lee-97,Carbotte-10}. 
Presumably these low temperature results indicate that we 
are not correctly evaluating the transverse Fermi velocity 
$v_{\Delta}$ and/or the longitudinal one $v_{f}$ therefore 
the proper linearity coefficient $\alpha=[2\ln{(2)}/\pi]
v_{f}/v_{\Delta}$ \cite{Lee-97} is missing. 
We also suspect that in our computations we might apparently 
overestimate the role of antinodal areas, where the majority 
of bosons is effectively gathered for $T \rightarrow 0$ 
(see figure 4 in the Ref.\ \cite{Ranninger-10}). 
This artificial low-temperature dependence of $n_{s}(T)$ 
needs a more careful investigation. 

Summarizing this section, we have obtained the superfluid
density $n_{s}(T)$ which clearly indicates a fairly broad 
temperature regime $T_{c}<T<T^{*}$ of the  Meissner 
rigidity appearing due to the superconducting fluctuations. 
Such fragile diamagnetism originates solely from the 
non-condensed preformed pairs, as has been previously 
suggested by several authors \cite{May-59,Zhu-94,Koh-03}. 
We have evaluated $n_{s}(T)$ by means of the new nonperturbative 
method and determined the current-current response function 
(\ref{response_BF}) solving iteratively the set of flow 
equations. Our results seem to qualitatively capture 
the experimental data of N.P.\ Ong group  \cite{Ong-10} 
ond other recent measurements \cite{Keren-11}.

\section{Summary}

We have addressed the linear response of the electron 
system with the pairing instabilities using nonperturbative 
framework of the continuous unitary transformation technique 
\cite{Wegner-94}. For the case of the Bose-Einstein condensed 
pairs we have analytically derived the BCS result 
(\ref{BCS_result}), which in the static and long wavelength 
limit accounts for the Meissner effect. We have next 
extended such treatment onto the mixture of the 
non-condensed (preformed) pairs interacting with the mobile 
electrons through the charge-exchange Andreev scattering. 
We have determined the contributions (see Fig.\ \ref{plot1}) 
to the response function (\ref{response_BF}), where the 
vertices are expressed by the corresponding flow equations 
(\ref{flow_A}-\ref{flow_F}). 

The central result (\ref{response_BF}) of our study generalizes 
the BCS current-current response function \cite{Fetter_book} 
taking into account the residual diamagnetic effects originating 
from the finite-momentum preformed pairs. Such effects are studied 
here in an alternative way than the perturbative corrections 
due to the Aslamasov-Larkin and Maki-Thompson diagrams 
\cite{Larkin-Varlamov}. In our approach the fluctations
enter the current-current response function through the 
convolution of one boson and two fermion propagators  
(see figure 1) instead of the higher order convolutions
typical for the standard  diagrammatic study. In the 
present formalism the influence of fluctuations affects
the vertex functions which have to be determined from 
the asymptotic solution of the flow equations 
(\ref{flow_A}-\ref{flow_F}).

In the static, long wavelength limit we find 
a clear evidence for the pronounced  diamagnetic 
contribution, which might be relevant 
to the experimental data obtained for the underdoped 
cuprate materials in the lower part of the pseudogap 
state \cite{Ong-10,Iye-10,Bernardi-10}. Our study is
consistent with the recent Quantum Monte Carlo results 
for the same model \cite{Troyer-11}. In both cases the 
residual diamagnetism originates from the preformed 
pairs whose mobility considerably increases below 
$T^{*}$. Similar ideas concerning the non-condensed 
pairs have been emphasized by several other authors 
\cite{Castellani-11,May-59,Zhu-94,Koh-03,
Tesanovic-08,Pieri-09,Kivelson-10}.

In order to see a more specific relation of this 
treatment to the cuprate oxides one should solve 
numerically the set of flow equations (\ref{flow_A}
-\ref{flow_F}) for the realistic model with nearest 
neighbor and next nearest neighbor hopping integrals. 
Another issue (not addressed here) concerns the doping 
effects which should affect the Fermi surface topology 
and influence the populations of the fermions and 
the preformed pairs \cite{Troyer-11,Domanski-01}. 
It would be also worthwhile to solve the flow equations 
(\ref{flow_A}-\ref{flow_F}) fully selfconsistently
and investigate the electrodynamic properties using 
the response function $K_{\alpha,\beta}({\bf q},
\omega)$, which generalizes the standard BCS result. 

We hope that the present formulation of the linear 
response theory by means of the flow equation method 
could stimulate further studies of the many-body effects 
in various subdisciplines of physics.

\begin{acknowledgments}
We acknowledge useful discussions with J.\ Ranninger and
K.I.\ Wysoki\'nski.  Moreover, T.D. wants to thank 
for a hospitality of the Neel Institute (CNRS, Grenoble), 
where the initial part of this study has been done. This 
project is supported by the Polish Ministry of Science 
and Education under the grant no.\ NN202187833.
\end{acknowledgments}

\begin{appendix}
\section{Methodology of the flow equations}
\label{A}

We give here a brief outline of the continuous canonical 
transformation for arbitrary Hamiltonian of the following 
structure
\begin{equation}
\hat{H} = \hat{H}_{0} + \hat{H}_{1} ,
\label{H0_H1}
\end{equation}
where $\hat{H}_{0}$ denotes the diagonal part (for 
instance it can be the kinetic energy of particles) 
and $\hat{H}_{1}$ stands for the off-diagonal term (i.e.\ 
interactions or any perturbations). Upon continuously 
transforming the Hamiltonian $H(l)=U^{\dagger}(l)HU(l)$ 
the $l$-dependence ({\em flow}) is governed according to 
the following differential equation \cite{Wegner-94}
\begin{eqnarray}
\frac{d\hat{H}(l)}{dl} = [ \hat{\eta}(l),\hat{H}(l) ]
\label{general}
\end{eqnarray} 
where the generating operator $\hat{\eta}(l) \equiv 
\frac{d \hat{{\cal{U}}}(l)}{dl} \hat{{\cal{U}}}^{-1}(l)$. 

Choice of $\eta(l)$ is usually dictated by the specific 
physical situation. One of the possibilities, suggested
by Wegner \cite{Wegner-94}, is 
\begin{equation}
\hat{\eta}(l) = \left[ \hat{H}_{0}(l), \hat{H}_{1}(l) \right] .
\label{wegner}
\end{equation}
It has been proved that (\ref{wegner}) guarantees 
\begin{equation}
\lim_{l\rightarrow \infty} \hat{H}_{1}(l) = 0
\label{limit}
\end{equation}
provided that no degeneracies are encountered. Several 
alternative proposals for $\hat{\eta}(l)$ capable to deal 
with the divergences \cite{Wilson-94}, the degenerate 
states \cite{Mielke-98} or various other advantages has
been discussed in the monograph \cite{Kehrein_book}.

To carry out the statistical averages of the
observables 
\begin{eqnarray}
\langle \hat{O} \rangle_{\hat{H}} = \mbox{Tr} 
\left\{ e^{-\beta \hat{H}} \hat{O} \right\}/\mbox{Tr} 
\left\{ e^{-\beta \hat{H}}\right\} 
\label{thermal_av}
\end{eqnarray}
(where $\beta^{-1}\!=\!k_{B}T$) it is convenient to use 
the invariance of trace on the unitary transformations  
\begin{eqnarray}
\mbox{Tr} \left\{  e^{-\beta \hat{H}} \hat{O} \right\} 
= \mbox{Tr} \left\{  e^{-\beta \hat{H}(l)} \hat{O}(l) 
\right\} ,
\label{trace}
\end{eqnarray}
where 
$\hat{O}(l)=\hat{{\cal{U}}}(l) \hat{O} \hat{{\cal{U}}}^{-1}(l)$. 
The $l$-dependence of $\hat{O}(l)$ is imposed 
through the flow equation  \cite{Wegner-94} 
\begin{eqnarray}
\frac{d\hat{O}(l)}{dl} = [ \hat{\eta}(l),\hat{O}(l) ] 
\label{O_flow}
\end{eqnarray}   
similar to (\ref{general}) for $\hat{H}(l)$.
Since the Hamiltonian  $\hat{H}(l)$ becomes diagonal for 
$l\rightarrow \infty$ the easiest way to compute the 
trace (\ref{trace}) is with respect to $\hat{H}(\infty)$. 
This however requires, that simultaneously with the 
continuous diagonalization of the Hamiltonian one has 
to analyze the {\em flow} of other  physical observables 
$\hat{O} \rightarrow \hat{O}(l) \rightarrow \hat{O}
(\infty)$.

\section{Effective quasiparticles above $T_{c}$}
\label{B}

To determine the single particle excitation spectrum for the 
model (\ref{BF}) we have to transform the individual operators  
$\hat{c}_{{\bf k}\sigma}^{(\dagger)}(l)\!\equiv\! \hat{U}(l)
\hat{c}_{{\bf k}\sigma}^{(\dagger)} \hat{U}^{-1}(l)$ which 
is a bit tricky because $\hat{U}(l)$ is not known explicitly. 
Following the scheme outlined in section II.B and using the 
operator $\hat{\eta}(l)$ chosen in the form (\ref{eta_BF}) 
we deduce the following ansatz for the fermion operators 
\cite{Domanski-03}
\begin{eqnarray}
&& \hat{c}_{{\bf k}\uparrow}(l)  =
u_{\bf k}(l) \; \hat{c}_{{\bf k}\uparrow}  + 
v_{\bf k}(l) \; \hat{c}_{-{\bf k}\downarrow}^{\dagger}
\label{c_Ansatz} \\ &&+
 \frac{1}{\sqrt{N}} \! \sum_{{\bf q} \neq {\bf 0}} \left[
u_{{\bf k},{\bf q}}(l) \; \hat{b}_{\bf q}^{\dagger}
\hat{c}_{{\bf q}+{\bf k}\uparrow}  + 
v_{{\bf k},{\bf q}}(l) \; \hat{b}_{\bf q} 
\hat{c}_{{\bf q}-{\bf k}\downarrow}^{\dagger}
\right] ,  \nonumber  \\
&& \hat{c}_{-{\bf k}\downarrow}^{\dagger}(l) = 
- v_{\bf k}^{*}(l) \; \hat{c}_{{\bf k}\uparrow} +
u_{\bf k}^{*}(l) \; \hat{c}_{-{\bf k}\downarrow}^{\dagger} 
\label{cbis_Ansatz} \\ &&
\frac{1}{\sqrt{N}} \! \sum_{{\bf q} \neq {\bf 0}} \left[  
- v_{{\bf k},{\bf q}}^{*}(l) \; \hat{b}_{\bf q}^{\dagger}
\hat{c}_{{\bf q}+{\bf k}\uparrow} +
u_{{\bf k},{\bf q}}^{*}(l) \; \hat{b}_{\bf q}
\hat{c}_{{\bf q}-{\bf k}\downarrow}^{\dagger}
\right] \nonumber , 
\end{eqnarray}
where $u_{\bf k}(0)=1$ and the other coefficients are vanishing
at $l\!=\!0$. These $l$-dependent coefficients can be derived 
from the equation (\ref{O_flow}) for $\hat{c}_{{\bf k}\uparrow}
(l)$ and $\hat{c}_{-{\bf k}\downarrow}^{\dagger}(l)$ operators.

The corresponding set of flow equations reads \cite{Domanski-03}
\begin{eqnarray}
\frac{d u_{\bf k}(l)}{dl} & = &  \sqrt{n_{{\bf q}\!=\!{\bf 0}}^{B}} 
\; \alpha_{-{\bf k},{\bf k}}(l) \; v_{\bf k}(l) 
\label{P_flow} \\ 
&+& \frac{1}{N} \sum_{{\bf q}\neq{\bf 0}} 
\alpha_{{\bf q}-{\bf k},{\bf k}}(l) \left( 
n_{\bf q}^{B} + n_{{\bf q}-{\bf k}\downarrow}^{F}
\right) v_{{\bf k},{\bf q}}(l), 
\nonumber \\ 
\frac{d v_{\bf k}(l)}{dl} & = & - \;
\sqrt{n_{{\bf q}\!=\!{\bf 0}}^{B}} \; \alpha_{{\bf k},{\bf k}}(l)
\; u_{\bf k}(l)
\label{R_flow}\\ &-& \frac{1}{N} \sum_{{\bf q}\neq{\bf 0}}
\alpha_{{\bf k},{\bf q}+{\bf k}}(l) \left(
n_{\bf q}^{B} + n_{{\bf q}+{\bf k}\uparrow}^{F}
\right) u_{{\bf k},{\bf q}}(l), \nonumber  \\ 
\frac{d  u_{{\bf k},{\bf q}}(l)}{dl} & = &
\alpha_{{\bf q}-{\bf k},{\bf k}}(l) \; v_{\bf k}(l) ,
\label{p_flow} \\
\frac{d  v_{{\bf k},{\bf q}}(l)}{dl} & = & - \;
\alpha_{{\bf k},{\bf q}+{\bf k}}(l)  u_{\bf k}(l) .
\label{r_flow}
\end{eqnarray}
They are additionally coupled to the flow equations 
(\ref{flow_g}-\ref{flow_E}) because of the terms
$\alpha_{{\bf k},{\bf k}'}(l)$. If one neglects the 
finite momentum boson states [when $u_{{\bf k},{\bf q}}
(l)=0=v_{{\bf k},{\bf q}}(l)$] these equations can be 
solved analytically \cite{Domanski-03}, reproducing 
the standard BCS case discussed in section II.B. The
case of ${\bf q}\neq{\bf 0}$ bosons is more cumbersome. 
We have previously studied such problem numerically 
\cite{Domanski-01,Domanski-03}, in particular 
considering also the 2-dimensional square lattice 
\cite{Ranninger-10} with the tight-binding dispersion 
$\xi_{\bf k}(0)=-2t \left[\mbox{\rm cos}(k_{x}a)+
\mbox{\rm cos}(k_{y}a)\right]-4t' \mbox{\rm cos}
(k_{x}a) \mbox{\rm cos}(k_{y}a)\!-\!\mu$ assuming 
the initial discrete energy $E_{\bf q}(0)\!=\!E_{0}$ 
and fixing the total charge concentration $2\sum_
{\bf q}n_{\bf q}^{B} + \sum_{\bf k} \left( n_{{\bf k} 
\uparrow}^{F}+n_{{\bf k}\downarrow}^{F} \right)$.
 
One of the valuable results obtained from such 
formalism concerned the pseudogap regime. Since 
above $T>T_{c}$ the condensate fraction is absent 
we can notice that (\ref{R_flow}-\ref{p_flow}) 
imply $v_{\bf k}(l)\!=\!0\!=\!u_{{\bf k},{\bf q}}(l)$. 
In other words the parametrization (\ref{c_Ansatz},
\ref{cbis_Ansatz}) simplify then to
\begin{eqnarray}
\hat{c}_{{\bf k}\uparrow}(l) = u_{\bf k}(l) \;
\hat{c}_{{\bf k}\uparrow} \;+ \; \frac{1}{\sqrt{N}} 
\sum_{{\bf q}\neq{\bf 0}} v_{{\bf k},{\bf q}}(l) \; 
\hat{b}_{\bf q} \hat{c}_{{\bf q}-{\bf k}
\downarrow}^{\dagger} 
\label{c_up} \\
\hat{c}_{-{\bf k}\downarrow}^{\dagger}(l) =
u_{\bf k}^{*}(l) \; \hat{c}_{-{\bf k}\downarrow}^{\dagger} 
- \frac{1}{\sqrt{N}} \sum_{{\bf q}\neq{\bf 0}}
v_{{\bf k},{\bf q}}^{*}(l) \; \hat{b}_{\bf q}^{\dagger}
\hat{c}_{{\bf q}+{\bf k}\uparrow} ,
\label{c_down}
\end{eqnarray}
with the coefficients obeying
\begin{eqnarray}
&&\frac{d u_{\bf k}(l)}{dl} = 
\frac{1}{N} \sum_{{\bf q}\neq{\bf 0}} 
\alpha_{{\bf q}-{\bf k},{\bf k}}(l) \left( 
n_{\bf q}^{B} + n_{{\bf q}-{\bf k}\downarrow}^{F}
\right) v_{{\bf k},{\bf q}}(l), 
\nonumber \\ 
&&\frac{d  v_{{\bf k},{\bf q}}(l)}{dl} = - \;
\alpha_{{\bf k},{\bf q}+{\bf k}}(l)  u_{\bf k}(l) .
\nonumber
\end{eqnarray}

Under such circumstances thus find that the single 
particle spectral function $A({\bf k},\omega)\!=\!
-\pi^{-1}\mbox{\rm Im} G_{\sigma}({\bf k},\omega
+i0^{+})$ takes the following structure
\begin{eqnarray}
&& A({\bf k},\omega) = | \tilde{u}_{\bf k}|^{2} 
\delta \left( \omega\!-\!\tilde{\xi}_{\bf k} 
\right) \label{spectral}\\
&&+ \frac{1}{N} \sum_{{\bf q}\neq{\bf 0}} 
\left( n_{\bf q}^{B} + n_{{\bf q}-{\bf k}\downarrow}^{F} 
\right) | \tilde{v}_{{\bf k},{\bf q}} |^{2} \delta 
( \omega\!+\! \tilde{\xi}_{{\bf q}
\!-\!{\bf k}} \!-\!\tilde{E}_{\bf q}) .
\nonumber
\end{eqnarray}
The first part of (\ref{spectral}) describes the 
quasiparticle whose renormalized dispersion 
$\tilde{\xi}_{\bf k}$ in the temperature regime 
$T<T^{*}_{sc}$ is discontinuous at the chemical potential 
(signaling a depletion of the low energy states). 
The second part of Eq.\ (\ref{spectral}) 
contributes the shadow to the above mentioned 
quasiparticle. These contributions constitute 
the characteristic Bogoliubov-type excitation 
spectrum which has been indeed observed experimentally 
in the yttrium \cite{Argonne-08} and lanthanum
\cite{Villigen-08} compounds.

\end{appendix}

\end{document}